\documentclass[twocolumn]{aastex631}
\usepackage{amsfonts,amsmath,graphicx,url,hyperref,nicefrac}
\usepackage{amssymb, verbatim, subfigure, paralist, soul, comment, multirow}
\usepackage{natbib}

\usepackage{color}
\makeatletter
\newcommand{\footnoterule}{%
  \kern -3pt
  \hrule width 0.4\columnwidth height 0.4pt
  \kern 2.6pt
}
\makeatother

\shorttitle{AT2018hyz: A Delayed Off-Axis Jet Model} 
\shortauthors{Dastidar \& Duffell}

\begin{document}

\title{AT2018hyz: Predictions for a Sky Projection from a Delayed Off-Axis Jet}

\author[0009-0000-6548-6177]{Ranadeep G. Dastidar}
\affiliation{Department of Physics and Astronomy, Purdue University, 525 Northwestern Avenue, West Lafayette, IN 47907, USA}

\author[0000-0001-7626-9629]{Paul C. Duffell}
\affiliation{Department of Physics and Astronomy, Purdue University, 525 Northwestern Avenue, West Lafayette, IN 47907, USA}

\begin{abstract}

Initially discovered optically, Tidal Disruption Event (TDE) AT2018hyz was first detected in radio 972 days later. The continued monitoring of the event revealed a late-time (1370-2160 days) radio brightening. The source for this radio brightening has been associated with an afterglow originating from an off-axis jet or a delayed mildly relativistic outflow. In this work, we explored the conditions required for a delayed off-axis jet to explain the observed radio brightening. We find that the combined effect of different delay times ($t_{\rm{delay}}$) and observer angles ($\theta_{\rm{obs}}$) can produce radio afterglow, as observed for AT2018hyz. The further off-axis from the jet core an observer is, the shorter is the delay inferred between the optical disruption and launching of the jet. This presents a degeneracy between different $\theta_{\rm{obs}}$ and $t_{\rm{delay}}$. We estimate the sky projection of the jet for these different combinations to break this degeneracy. The spread in sky images ranges from $\sim$ 0.6 mas for a $30^{\circ}$ off-axis observer to  $\sim$ 0.3 mas for an observer perpendicular to the jet core. Future VLBI observations should be able resolve these structures. Thus a combined observation of the light curve and sky images for AT2018hyz (and other jetted TDEs) can be used to narrow down the observer angle and any delay in launching of the underlying jet.\\

\end{abstract}


\section{Introduction} \label{sec:intro}

A star wandering close enough to a supermassive black hole (SMBH) can be torn apart by its tidal forces; this is called a tidal disruption event (TDE). A significant fraction of the disrupted stellar material is accreted by the SMBH, producing bright transitory multi-band electromagnetic radiation flares (\cite{Rees+1988b, Komosa+2015}). While TDEs have been discovered in multiple wavebands, optical discoveries account for most of them (see \cite{Gezari+2021} for a detailed review), with follow-up in other bands. TDE emissions in the X-ray, UV, and optical bands reveal the underlying fallback rates and accretion of the disrupted star onto the SMBH (\cite{Halpern+2004, Ladato+2011, Guillochon+2013, Law-Smith+2020, Ryu+2020}). A few $\gamma$-ray detections, followed up in X-ray and centimeter-to-millimeter wavelengths, have revealed relativistic jetted outflows from TDEs (\cite{Levan+2011, Burrows+2011, Zauderer+2011}). Radio observations have also hinted at the presence of collimated relativistic jets associated with TDEs (\cite{Bloom+2011, Giannios+2011, DeColle+2012}). Population studies have shown that nearly 10\% of the TDEs launch a jet (\cite{Bower+2013, Velzen+2013, Brown+2015}). Furthermore, a few TDEs observed in radio have been associated with non-relativistic outflows (\cite{Alexander+2016, Stein+2020, Cendes+2021}).

More recently, \cite{Cendes+2024} have reported late-time radio emissions from TDEs, hundreds of days since their optical disruption, despite earlier non-detections in nearly 40\% of all the optically detected TDEs. Multiple efforts have been made to model the physical origin of these delayed radio emissions. These include multiple episodes of energy injection (\cite{Goodwin+2025}), non-relativistic disruption outflow interacting with circumnuclear gaseous clouds (\cite{Mou+2022, Zhuang+2025}) or its interaction with a non-linear circumnuclear density (\cite{Matsumoto+2024}), and radio flares from stream-stream interactions (\cite{Goodwin+2023, Hajela+2025}). One of the leading theories is radio emission from off-axis sources (\cite{Zauderer+2011, Sfaradi+2024, Matsumoto+2023}), which becomes visible to the observer once the jet has decelerated sufficiently. These models account for the delay in emission to a delayed interaction time between the promptly ejected outflow from the tidal disruption and the circumnuclear material. In contrast, after the disruption, a reduced mass-fallback rate due to circularization can slow the accretion process, thereby delaying the launch of the outflow from the TDE (\cite{Hayasaki+2021}). The time taken for circularization can launch a delayed sub-relativistic or mildly relativistic outflow over timescales of hundreds to thousands of days post-disruption (\cite{Horesh+2021, Cendes+2022, Zhang+2024, Wu+2026}). In many such TDEs, both the delayed mildly relativistic outflow or off-axis jet scenarios can be incorporated to model the observations (\cite{Alexander+2026, Cendes+2026}), while in some cases neither model fit (\cite{Christy+2024}). Despite many such efforts, a coherent theory for delayed radio emissions in TDEs does not yet exist.

\cite{Cendes+2022} reported a rapidly rising radio emission from a tidal disruption event, AT2018hyz, 972 days after its optical discovery (\cite{Horesh+2018, Gomez+2020}). The radio emissions were initially interpreted to originate from a mildly relativistic outflow launched about 700-750 days after the initial disruption (\cite{Cendes+2022}). Later studies have also proposed an extremely off-axis relativistic jet ($\sim 70^{\circ}-80^{\circ}$) as a possible model (\cite{Matsumoto+2023, Sato+2024, Sfaradi+2024}). A combined discussion and contrast of the two interpretations, incorporating more recent observations, can be found in \cite{Cendes+2026, Matsumoto+2025}. In other similar TDEs, it has been found that less powerful jets can escape the system after aligning with the SMBH spin, leading to a delayed jet launch in some TDEs (\cite{Teboul+2023, Goley+2025}). In either scenario, radio imaging with very long baseline interferometry can be used to distinguish between a mildly relativistic quasi-spherical outflow and a relativistic jet (\cite{Matsumoto+2023, Goley+2025, Matsumoto+2025}). Different centroid spreads and apparent motions in VLBI radio images are expected for the two origin scenarios for AT2018hyz (and other similar astronomical transients).

In this context, we explore AT2018hyz as a delayed jet viewed off-axis. In section \ref{sec:TheoMod}, we introduce the jet and afterglow models used for this study. This section also introduces the hydrodynamic and afterglow codes used. This is followed by a discussion of the techniques and MCMC strategies used to analyze the results in section \ref{sec:Analysis}. We then list the results that satisfactorily explain the observed radio brightening in section \ref{sec:Results}. The implications for these results and details for the sky images for the event are discussed in section \ref{sec:Discussion} and are followed by our conclusions for the study.

\section{Methods} \label{sec:TheoMod}

In this work, we model a relativistic jetted outflow confined within a jet opening angle $\theta_0$. The jet is launched with an isotropic equivalent energy $E_{\rm{iso}}$ and travels through a constant density interstellar medium (ISM) with density $n_{\rm{ISM}}$. The afterglow light curves from synchrotron emission due to the jet-ISM interaction are then generated and compared with the observations reported in \cite{Cendes+2026} for AT2018hyz in radio frequencies. We explore the conditions required for a delayed jet launch, observed off-axis, to account for the late-rising radio emissions in AT2018hyz. A Monte Carlo Markov Chain (MCMC) analysis was used to identify the set of jet parameters ($E_{\rm{iso}}$, $n_{\rm{ISM}}$, and the delay time for jet launch $t_d$) for different off-axis observer angles ($\theta_{\rm{obs}}$) that can explain the observations. Against each satisfactory model, we then estimate the apparent sky image and centroid motion projected by the jet interaction. We predict the centroid size and motion at 3000 days post-disruption (end of 2026) at one-year intervals to distinguish among the different models. In this section, we discuss the models and methods used for this analysis.

\subsection{The Jet Model}

We chose a double-sided boosted fireball to model our structured jet \citep{Wu+2018, Dastidar+2024, Schroeder+2025}. A boosted fireball is a two-parameter structured jet model that generates a family of outflows after they have expanded orders of magnitude larger than the disruption scale \citep{Duffell+2013}. A fireball is launched with a fluid frame Lorentz factor $\eta_0$ and then boosted along the jet axis with an observer frame Lorentz factor $\gamma_B$. In the fluid frame, the fireball expands with the Lorentz factor $\eta_0$ given by the total internal energy E and the mass M, where $\eta_0 \sim E/M$. Due to the $\gamma_B$ boost in the observer frame, this fireball is then naturally confined with an opening angle $\theta_0 \approx 1/\Gamma_B$. The total jet energy in each fireball in the observer frame thus becomes $E_0 \sim \gamma_B\eta_0E$. Thus, we end with a model for a structured jet with total energy $E_0$ and opening angle $\theta_0$. Two independent fireballs are launched, boosted opposite to each other, to simulate both the jet and the counter jet. Thus, the isotropic equivalent energy ($E_{\rm{iso}}$) is related to the jet energy $E_0$, as

\begin{equation}
\label{eq:Eiso}
    2E_0 \simeq 4\pi(\theta_0)^2\frac{dE}{d\Omega} \sim \frac{E_{iso}}{\gamma_B^2}
\end{equation}

The boosted fireball model for the jet naturally attains a structured jet form where the energy distribution is given by:

\begin{equation}
\label{Eq:dEdOmega}
    \frac{dE}{d\Omega} = \frac{E_{\rm{iso}}}{4\pi}\left(\frac{\gamma_{\rm{max}}(\theta)}{\gamma_{\rm{max}}(0)}\right)
\end{equation}

Where, $$\gamma_{\rm{max}}(\theta) = \gamma_B\frac{\eta_0 + v_B\rm{cos}\theta\sqrt{\eta_0^2v_0^2 - \gamma_B^2v_B^2sin^2\theta}}{1 + \gamma_B^2v_B^2sin^2\theta}$$ (see \cite{Duffell+2013} for detailed derivation). Once initialized for a given set of ($\eta_0$, $\gamma_B$), the jet structure then evolves following special relativistic hydrodynamic conservation equations, hence modeling the jet spread naturally as the jet evolves \citep{Wang+2024}. Based on observational findings for AT2018hyz \citep{Cendes+2026, Matsumoto+2023, Matsumoto+2025}, we have further modeled the interstellar medium jet expands into as a constant density $\rho_{\rm{ISM}} = n_{\rm{ISM}}$.

\begin{figure}
\centering
\includegraphics[width=\linewidth]{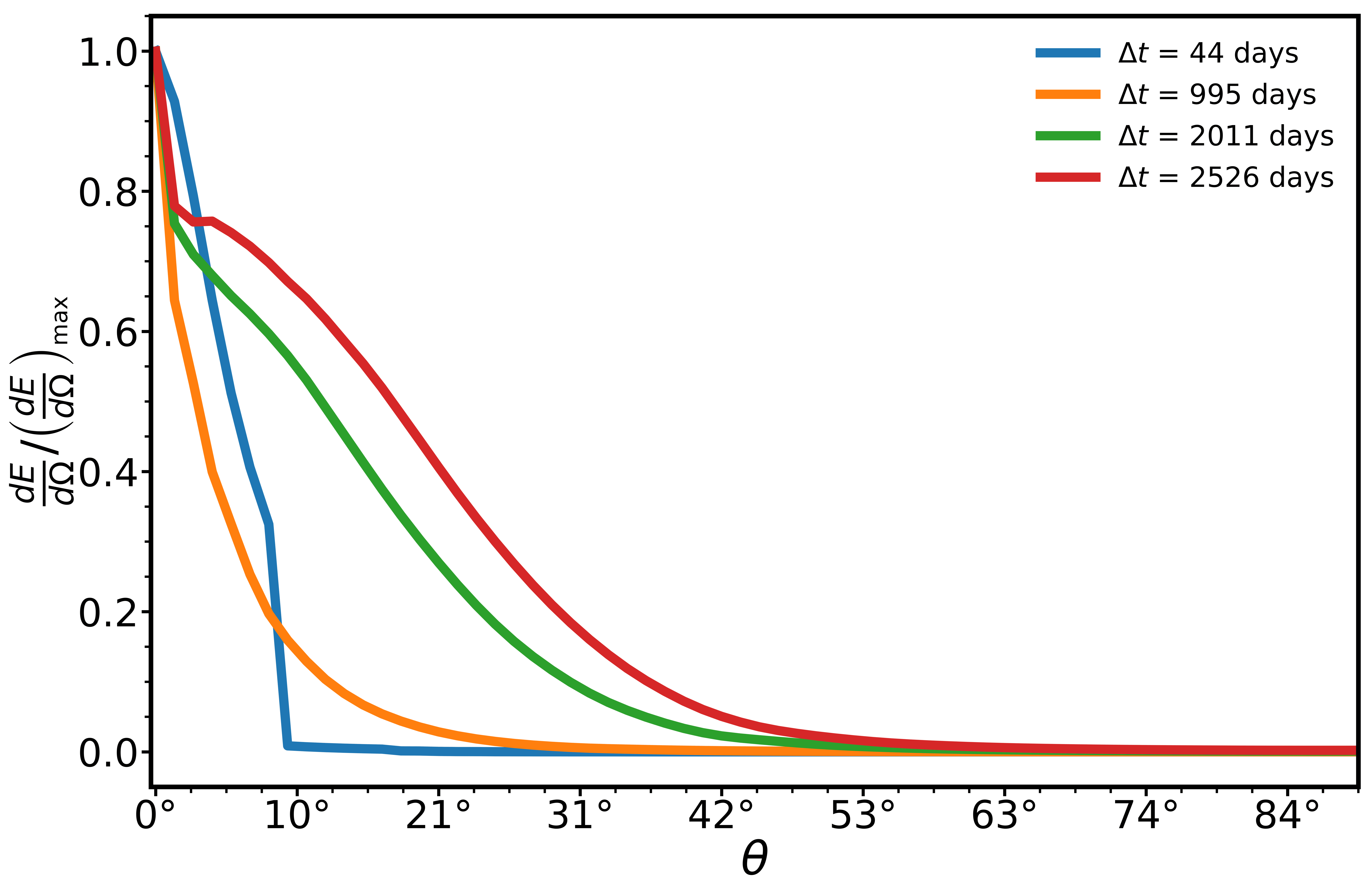}
\caption{\footnotesize{This shows the angular energy distribution for the jet as it expands and spreads. We initiate the jet as a fireball with the energy confined within an opening angle of $10^{\circ}$. With time, the profile broadens as the jet spreads. The energy distribution has been normalized with respect to its on-axis value.}}
\label{fig:lc_fit_multi_Hz}
\end{figure}

The hydrodynamical evolution and ISM interaction of the jet is carried out using \textit{The Jet Code} \citep{Duffell+2018}. \textit{The Jet Code} is a 2-dimensional moving-mesh relativistic hydrodynamics solver. The moving mesh resolves the thin shock width for a very high Lorentz factor extremely well. The simulations are carried out in a 2-dimensional axisymmetric $r-\theta$ plane. The system is initialized with a radial resolution, $\delta R/R = 1/512 \approx 2\times10^{-3}$ and angular resolution, $\delta\theta/\theta = 1/256 \approx 4\times10^{-3}$. The simulation begins much before the fireball enters the Blandform-McKee self similar phase, that is $t_{\rm{min}} = 0.06t_{\rm{BM}}$, and is carried out until 30 times the Blandford-McKee time scale, that is $t_{\rm{max}} = 30t_{\rm{BM}}$. Where, the Blandford-McKee time scale is given by $t_{\rm{BM}} \sim (E_{\rm{iso}}/(\rho_0c^5\Gamma^2))^{1/3}$. In actual time, the simulation duration translates to an expansion from a few days to after 50 years since the jet launch. This ensures that the jet evolution is tracked from before the deceleration phase through the Sedov-Taylor subrelativistic phase. The system is stratified according to the density profile given by Eq. \ref{eq:rho_ini}, and a homologous expansion of the ejecta is assumed.

 \begin{equation}\label{eq:rho_ini}
     \rho(r) = \rho_{\rm{max}}\left(\frac{1-R_{\rm{sh}}/t}{1-r/t}\right)
 \end{equation}

\begin{eqnarray}\label{eq:vel_ini}
      v(r,t) = 
  \begin{cases}
      r/t & \text{if $v < R_{\rm{sh}}$}\\
      0 & \text{otherwise}\\
      
  \end{cases}
\end{eqnarray}

Here $\rho_{\rm{max}}\sim E/4\pi c^3t^3$ is the maximum blast wave density in the shock and follows from $E=4\pi\rho_{\rm{max}}(ct)^3c^2$. The shock radius ($R_{\rm{sh}}$) is given by $R_{\rm{sh}} = t(1-1/2\eta_0^2)$. The initial pressure is set to a very low value ($10^{-5}\rho$), and an adiabatic equation of state with a constant adiabatic index of 4/3 is used for relativistic fluids.

\subsection{Afterglow Physics}

For the jet evolution discussed above, we then model the afterglow emission from the jet-ISM interaction. The afterglow emission is modeled as synchrotron radiation originating from the jet's forward shock as it accumulates material from the surrounding medium. Afterglow calculations are performed using the \textit{Firefly} code \citep{Dastidar+2024}. The \textit{Firefly} code is based on the broken power-law standard afterglow model \citep{Sari+1998, Granot+2002, Eerten+2013}. The code takes the hydrodynamically evolved system as its input and sums the synchrotron emissivity from each hydrodynamic cell, weighted by their Doppler boosting for a given observer. This helps model afterglows in a way that is agnostic to any presumed jet structure and captures the effects of jet spread on afterglow emission. For systems evolved in 2 dimensions, \textit{Firefly} first converts the hydro data into 3 dimensions, assuming the entire domain is axi-symmetric around the jet axis. This is not required for on-axis observations but is necessary for off-axis observers. The \textit{Firefly} code is publicly available on GitHub and archived on Zenodo \citep{Dastidar+2026}.

\subsubsection*{Light Curves}
The afterglow calculations are carried out as follows. For each cell on the hydrodynamic output (for a given time step), the emissivity in the local fluid frame ($\epsilon'_{\nu'}$) is calculated as:

\begin{equation}
\label{eq:epsE}
    \epsilon'_{\nu'} = \epsilon'_p\times
      \begin{cases}
      (\nu'/\nu'_m)^{1/3} & \text{if $\nu' < \nu'_m < \nu'_c$}\\
      (\nu'/\nu'_m)^{-(p-1)/2} & \text{if $\nu'_m < \nu' < \nu'_c$}\\
      (\nu'_c/\nu'_m)^{-(p-1)/2}(\nu'/\nu'_c)^{-p/2} & \text{if $\nu'_m < \nu'_c < \nu'$}\\
      (\nu'/\nu'_c)^{1/3} & \text{if $\nu' < \nu'_c < \nu'_m$}\\
      (\nu'/\nu'_c)^{-1/2} & \text{if $\nu'_c < \nu' < \nu'_m$}\\
      (\nu'_m/\nu'_c)^{-1/2}(\nu'/\nu'_m)^{-p/2} & \text{if $\nu'_c < \nu'_m < \nu'$}\\
      \end{cases}
\end{equation}

where,

\begin{equation}
    \epsilon'_p = \frac{\sqrt{3}e^3B'fn'}{m_ec^2}
\end{equation}

Break frequencies for the broken power law are obtained by the relation $\nu_i' = \frac{3eB'\gamma'^2_e}{4\pi m_e c}$, by substituting for the minimum Lorentz factor for electron acceleration, $\gamma'_m$, and the Lorentz factor for electron cooling, $\gamma'_c$. Assuming a power law electron distribution for accelerated electrons as $N(\gamma') \propto \gamma'^{-p}$, yields $$\gamma_m' = \frac{p-2}{p-1}\frac{\epsilon_em_p}{fm_e}(\Gamma-1),$$ for local fluid Lorentz factor $\Gamma$ and fraction of kinetic to total thermal energy $\epsilon_e$, \textit{f} is the fraction of electrons radiating from the total available population. We find that variations in \textit{f} have no significant effect on the light curves, and it is fixed to be unity from here onward. The cooling break is estimated on the basis of a global cooling time scale $t'$ ($t = \gamma t'$), and can be solved for as $$\gamma'_c = \frac{6\pi m_e\gamma c}{\sigma_T B'^2t}$$. Where the magnetic field strength $B'$ is approximated by the parameter $\epsilon_B$ as the fraction of total magnetic energy to total thermal energy ($e'_{\rm{th}}$). It is obtained as, $$\epsilon_B = \frac{B'^2}{8\pi e'_{\rm{th}}}$$.

Finally, the afterglow flux is calculated by integrating the emissivity over the entire jet evolution, with respect to an observer placed at $\hat{n}$, on an axisymmetric \textit{x-z} plane, where the jet moves along the $+\hat{z}'$ direction. The unit vector $\hat{n}$ subtends an angle $\theta_{\rm{obs}}$ with respect to the $\hat{z}$ axis. This gives the observed flux ($F_{\nu}(\nu_{\rm{obs}},t_{\rm{obs}})$) as:

\begin{equation}
    F_{\nu}(\nu_{\rm{obs}},t_{\rm{obs}}) = \frac{1+z}{4\pi D_L^2}\int\epsilon'_{\nu'}\times\delta^2\times dV
\end{equation}.

Where the primed quantities are calculated in the fluid rest frame, $D_L$ is the luminosity corresponding to the gravitational redshift of the source $z$. The Doppler correction ($\delta$) and observer frame frequency ($\nu_{\rm{obs}}$), and observer time ($t_{\rm{obs}}$) can be calculated as:

\begin{equation}
\label{eq:Doppler}
    \delta  = \Gamma^{-1}(1-\Vec{v}\cdot\Hat{n})^{-1}
\end{equation}

\begin{equation}
    \nu_{\rm{obs}} = \frac{\delta}{1+z}\nu'
\end{equation}

\begin{equation}
    \frac{t_{\rm{obs}}}{(1+z)} = t'-\frac{\Vec{r}'\cdot\Hat{n}}{c}
\end{equation}

\subsubsection*{Sky Projection}

\begin{figure}
\centering
\includegraphics[width=0.75\linewidth]{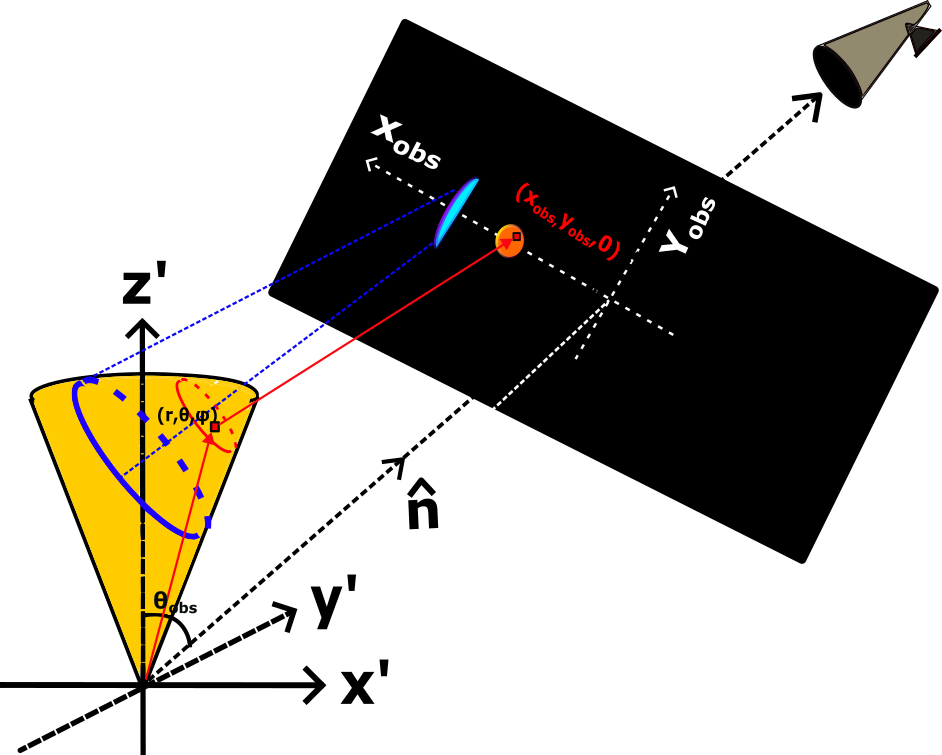}
\caption{\footnotesize{Pictorial description of 3D jet projection on a 2D sky projections. Each cell ($r,\theta,\phi$) on the equal arrival time surface for a user-provided $t_{\rm{obs}}$ and frequency ($\nu_{\rm{obs}}$) is uniquely reflected on the observer plane ($x_{\rm{obs}}, y_{\rm{obs}},0$) using Eq. \ref{Eq:Sky_projection} to generate the observed sky image.}}
\label{fig:sky_projection_cartoon}
\end{figure}

The \textit{Firefly} code can also estimate the sky image projected by the jet interaction, with respect to the observer. Since this afterglow code is independent of the jet model and hence its evolution (both radial expansion and angular spread), the sky projection of the jet centroid captures the effects of the jet spread as well. Within the \textit{Firefly} code, the sky projection is obtained by reflecting the flux emitted from the jet, at a given observer time and frequency, on a 2D plane perpendicular to the observer position vector $\hat{n}$. The x-axis of this plane is constructed along the motion of the jet along this plane. Geometrically, this can be represented as,

\begin{equation}
    \hat{x}_{\rm{obs}} = (\hat{z}' - (\hat{z}'.\hat{n})\hat{n})/\sqrt{1-(\hat{z}'.\hat{n})^2}
\end{equation}

Then, by construction, the y-axis of this projection plane is perpendicular to ($\hat{x}_{\rm{obs}},\hat{n}$), and $z_{\rm{obs}} = 0$ always. Calculated explicitly for an observer placed at $\hat{n}$, which is ($\theta_{\rm{obs}},0$) is the $\theta-\phi$ plane, each hydrodynamic cell placed at ($r,\theta,\phi$) would project to ($x_{\rm{obs}},y_{\rm{obs}},0$) in the observer sky. Solving for the x-y coordinates, we get:

\begin{equation}
\begin{aligned}\label{Eq:Sky_projection}
    x_{\rm{obs}} = W_x &= r[cos(\theta)sin(\theta_{\rm{obs}}) - sin(\theta)cos(\phi)cos(\theta_{\rm{obs}})]\\
    y_{\rm{obs}} = W_y &= -r[sin(\theta)\sin(\phi)]
\end{aligned}
\end{equation}

For a source at redshift $z$, and luminosity distance $D_L$, the angular distance then becomes $w_i = W_i/(D_L\times(1+z)^2)$, where $W_i$ ($i \in {x,y}$) is the actual distance from the source. A pictorial description of the geometry is shown in Fig. \ref{fig:sky_projection_cartoon}. We then calculate the flux-weighted centroid location as:

\begin{equation} \label{Eq:Centroid_location}
    X_c = \frac{\int x_{\rm{obs}}F_{\nu}dx_{\rm{obs}}dy_{\rm{obs}}}{\int F_{\nu}dx_{\rm{obs}}dy_{\rm{obs}}}
\end{equation}

Once the centroid location is obtained over different observer time intervals, we find the apparent velocity ($v_{\rm{app}}$) observed for the jet centroid as:

\begin{equation}\label{Eq:vapp}
    v_{\rm{app}} = (D_L\times(1+z)^2)\Delta\theta_c/\Delta t_{\rm{obs}} 
\end{equation}

Where $\Delta\theta_c$ is the change in the centroid angular distance over the observed time interval $\Delta t_{\rm{obs}}$.

\section{Analysis} \label{sec:Analysis}

We perform a Monte Carlo Markov Chain (MCMC) analysis on our simulated afterglow light curves against the observed data for AT2018hyz (reported in \cite{Cendes+2026}). However, as discussed in the previous section, we generate our light curves using the \textit{Firefly} code directly from the hydrodynamic simulations. To capture the jet spread and model the exact light curves, we used the fully two-dimensional, highly resolved, axisymmetric hydrodynamic evolution of the jet. This consequently makes the afterglow light-curve calculations longer ($\sim 1$ hour) and inefficient for carrying out MCMC analysis. To work around this, we exploit the self-similarity of blast wave expansion during the Blandford-McKee phase to generate rapid light curves.

The self-similarity of the jet expansion translates into scalable light curves in isotropic equivalent energy ($E_{\rm{iso}}$) and ISM density ($n_{\rm{ISM}}$) \citep{Eerten+2024}. That is, for a given set of jet parameters ($\gamma_B,\eta_0$ in our models) and afterglow parameters (electron distribution power-law slope $p$, $\epsilon_e$, $\epsilon_B$, $\theta_{\rm{obs}}$, and $\nu_{\rm{obs}}$), a family of afterglow light curves for different $E_{\rm{iso}}$ and $n_{\rm{ISM}}$ can be generated by re-scaling a fiducial light curve ($F_0(t_{\rm{obs}})$ say with, $E_{\rm{iso},0}$ and $n_{\rm{ISM},0}$). Hence, the afterglow flux can be written as $F(t_{\rm{obs}}) \propto E_{\rm{iso}}^an_{\rm{ISM}}^b$. Thus, for a scaling, $E'_{\rm{iso}}=\kappa E_{\rm{iso},0}$ and $n'_{\rm{ISM}} = \lambda n_{\rm{ISM},0}$, we get the new flux ($F'(t'_{\rm{obs}})$) by the transformations:

\begin{equation} \label{Eq:scaling}
    \begin{aligned}
        F'(t'_{\rm{obs}}) &= F_0(t_{\rm{obs}})\left[\frac{E'_{\rm{iso}}}{E_{\rm{iso},0}}\right]^a\left[\frac{n'_{\rm{ISM}}}{n_{\rm{ISM},0}}\right]^b\\
        & = F_0(t_{\rm{obs}})\times \kappa^a\times \lambda^b\\
        t'_{\rm{obs}} &= t_{\rm{obs}}\left[\frac{E'_{\rm{iso}}}{E_{\rm{iso},0}}\right]^c\left[\frac{n'_{\rm{ISM}}}{n_{\rm{ISM},0}}\right]^d\\
        &= t_{\rm{obs}}\times \kappa^a\times \lambda^b\\
    \end{aligned}
\end{equation}

for some scaling factors {a,b,c, and d}. These scaling factors are sensitive to the other jet, afterglow, and observer parameters. \cite{Eerten+2024} lists the various scaling values analytically. In this work, we numerically determine these scalings using multiple simulated light curves from \textit{Firefly}. In this way, we preserve the resolution and the effects of jet spreading in the scaled light curve. This reduces the time to generate a light curve to less than a second, providing a basis for carrying out MCMC analysis with them.

We perform a three-parameter MCMC analysis of our light curves and the observed data. We chose the scalable equivalent isotropic energy ($E_{\rm{iso}}$), the ISM density ($n_{\rm{ISM}}$), and the delay time of the jet launch ($t_{\rm{delay}}$), that is, the time between disruption and jet launch, as free parameters for our random walker to explore. We assume an independent uniform prior distribution for each parameter. The MCMC is carried out with 32 walkers, each taking 5000 steps and discarding 1000 burn-in steps to ensure convergence. Increasing the steps or burn-in gives the same results. We used the readily available \textit{emcee} Python package to implement the MCMC on our scalable light curve models to report our parameter estimates within the $16^{th}$, $50^{th}$, and $84^{th}$ percentiles. We further calculate the Gelman-Rubin $\hat{R}$ statistic to assess convergence of our results.

\section{Results} \label{sec:Results}
To model our afterglow light curves, we explore the afterglow parameter space reported by \cite{Cendes+2026, Matsumoto+2025}. We consider a boosted fireball-structured jet mode with a jet opening angle of $\theta_j = 0.1$ rad, corresponding to $\gamma_B = 10$. The jet expands into a constant density interstellar medium, $\rho_{\rm{ISM}} \propto Ar^{0} = n_{\rm{ISM}}$. We explored a slew of models for the fireball intrinsic Lorentz factor, $\eta_0$, and found that $\eta_0 = 10$ fits the observed rising slope in the light curve for AT2018hyz. In addition, the micro-physics parameters in the standard afterglow model are held fixed at $\epsilon_e = 0.07$, and $\epsilon_B = 10^{-3}$. The electron distribution power-law slope 'p' (i.e. $N(\gamma) \propto \gamma^{-p}$) can be estimated from the multiband spectral data. However, as reported in \cite{Cendes+2026}, the multi-epoch spectral observations yield the p-value ranging from 1.77 to 2.58. Consequently, we generated multiple afterglow plots within the physically permissible range of $2.0 < p < 2.6$, and found that $p=2.05$ reproduced the observed rising slope in light curves. Holding these parameters ($\theta_j$, $\epsilon_e$, and $\epsilon_B$) fixed, with the source distance at redshift z = 0.0457, we generate light curves for an off-axis observer placed at $\theta_{\rm{obs}} = (10,20,30,...,80,90)$, and explore the three parameters $E_{\rm{iso}}$, $n_{\rm{ISM}}$, and the jet launch delay time ($t_{\rm{delay}}$) against each observer angle to explain the observed delayed radio light curves for AT2018hyz in the radio band (5 GHz, 9 GHz, and 20 GHz) using MCMC. Our findings are reported below.

\begin{table}[ht]
\centering
\begin{tabular}{c|cccc}
\hline
\multirow{2}{*}{$\theta_{\rm{obs}}$} & \multicolumn{4}{c}{Jet Parameters} \\
\cline{2-5}
 & $E_{\rm{iso,52}}$ (ergs) & $n_{\rm{ISM}}$ (cm$^{-3}$) & $t_{\rm{delay}}$ (days) & $\hat{R}$ \\
\hline
30$^{\circ}$ & 184.51 & 0.03 & 1072.82 & 1.0080\\
40$^{\circ}$ & 116.73 & 0.25 & 1044.69 & 1.0064\\
50$^{\circ}$ & 88.59 & 1.05 & 1010.41 & 1.0060\\
60$^{\circ}$ & 76.08 & 3.11 & 976.14 & 1.0045\\
70$^{\circ}$ & 61.80 & 7.69 & 948.61 & 1.0046\\
80$^{\circ}$ & 45.99 & 15.15 & 922.23 & 1.0031\\
90$^{\circ}$ & 34.08 & 20.18 & 902.82 & 1.0031\\
%30$^{\circ}$ & Value 1 & Value 2 & Value 3 \\
\hline
\end{tabular}
\caption{\footnotesize{Estimated values for the isotropic equivalent energy, ISM density, and delay time (for $\nu_{\rm{obs}} = 9$ GHz) in launching the jet against different observer angles using MCMC. These are the different combinations required to explain the observation for AT2018hyz. All other jet and afterglow parameters are the same for all $\theta_{\rm{obs}}$ and listed in the text. The last column shows the $\hat{R}$ statistics for confidence of convergence.}}
\label{tab:Thobs_Eiso_nism}
\end{table}

We find that light curves from observers with $\theta_{\rm{obs}} < 30^{\circ}$ cannot account for the observations. The MCMC does not converge to physical values for isotropic energy, ISM density, or any delay time that gives rise to the observed radio brightening. For further off-axis observers, that is, $\theta_{\rm{obs}} > 30^{\circ}$, the three-parameter MCMC converges to a unique set of values for each observer angle that can explain the radio brightening. For convergence, we accept only those parameters that have a Gelman-Rubin statistic $\hat{R} \approx 1$. Similar corner plots are generated for all the observer angles but are not shown in this paper. The MCMC results for each observer angle are shown in Table \ref{tab:Thobs_Eiso_nism}.

\begin{figure}
\centering
\includegraphics[width=\linewidth]{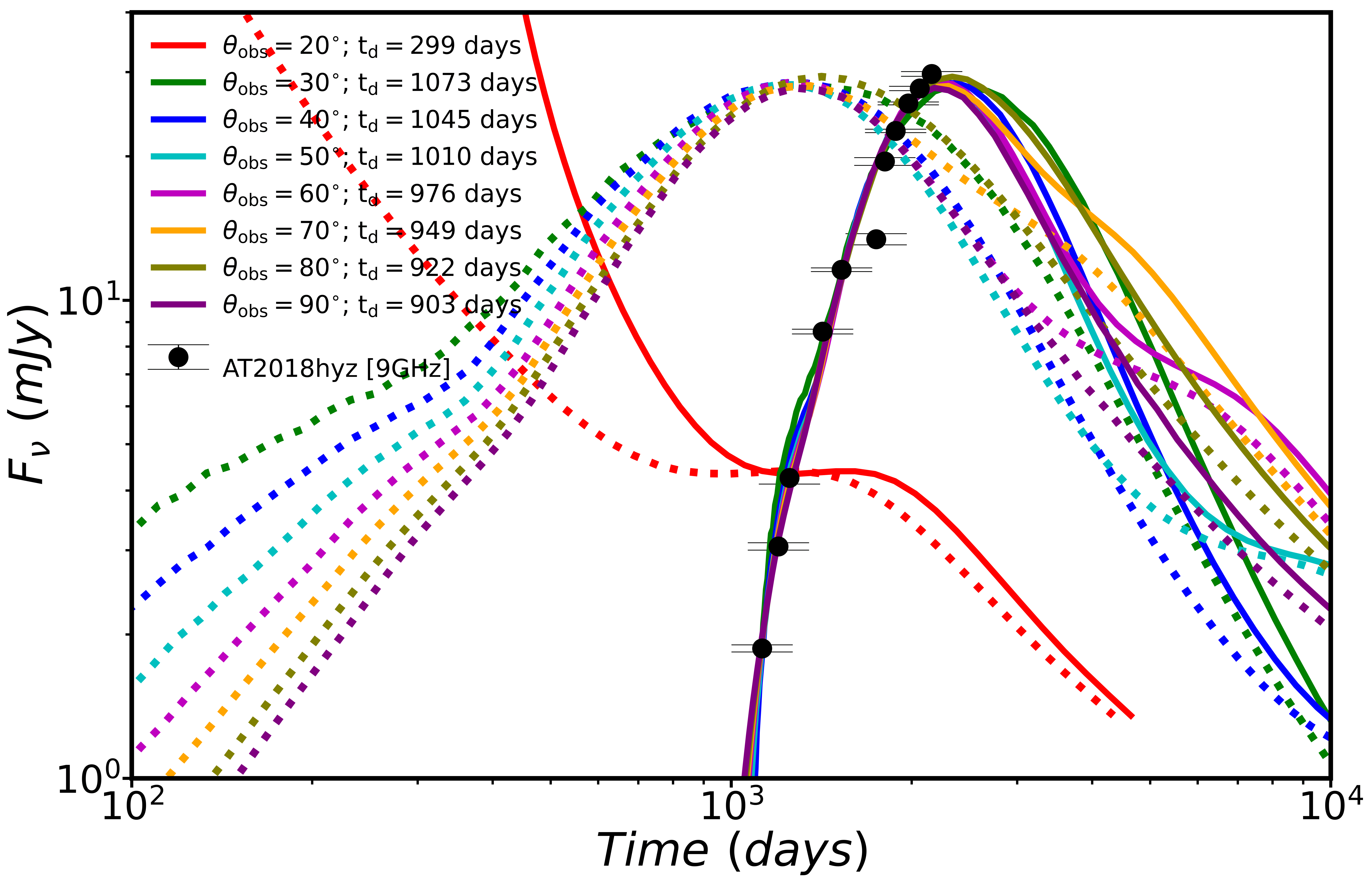}
\caption{\footnotesize{Light curves at 9 GHz for different off-axis observer angles, with (solid lines) and without (dashed) delay in jet launching. The observed data (\cite{Cendes+2026}) at 9 GHz are shown as solid black circles. A jet launched long ($\sim 1000$ days) after the disruption, observed closer to its axis ($> 30^{\circ}$), and a shorter delayed launch, observed further off-axis, can both explain the observed light curve very well. No amount of delay for the $20^{\circ}$ off-axis light curve (red) can explain the observations, and is shown for comparison.}}
\label{fig:lc_wo_w_delay}
\end{figure}

We then generate light curves using these parameters, and compare them against the observed data. Fig. \ref{fig:lc_wo_w_delay} shows a comparison of simulated light curves from the \textit{Firefly} code, overplotted with the 9 GHz radio data. For comparison, we have also shown the afterglow light curves (dotted lines) without any delay between the disruption and jet launching. We find that any observer angle beyond $30^{\circ}$ with appropriate $E_{\rm{iso}}$, $n_{\rm{ISM}}$, and $t_{\rm{delay}}$ satisfactorily explains the observed data. Accordingly, the 9 GHz emission peak should occur around 2300 days after the optical disruption. This peak-flux timescale varies very slightly with the observer's angle. However, the true variation in light curves between the different observer angles would only be evident at much later epochs.

MCMC for different observer frequencies (5 GHz, 9 GHz, and 20 GHz) converge to slightly different parameter sets. However, for the purposes of this study, we select the parameter values that best fit the 9 GHz data. The light curves thus generated at other frequencies with these parameters also explain the data within reasonable limits. This is shown in Fig. \ref{fig:lc_fit_multi_Hz}, where we plot the multi-band light curves in radio frequencies. The figure shows the observed data at 5 GHz, 9 GHz, and 20 GHz against the light curves for a delayed jet launched for off-axis observers at $30^{\circ}$ to $ 90^ {\circ}$ with different transparencies. The afterglow parameters used for these light curves are taken from Table \ref{tab:Thobs_Eiso_nism}, while keeping the other parameters fixed. All the light curves explain the observed data within reasonable margins.

\begin{figure}[h]
\centering
\includegraphics[width=\linewidth]{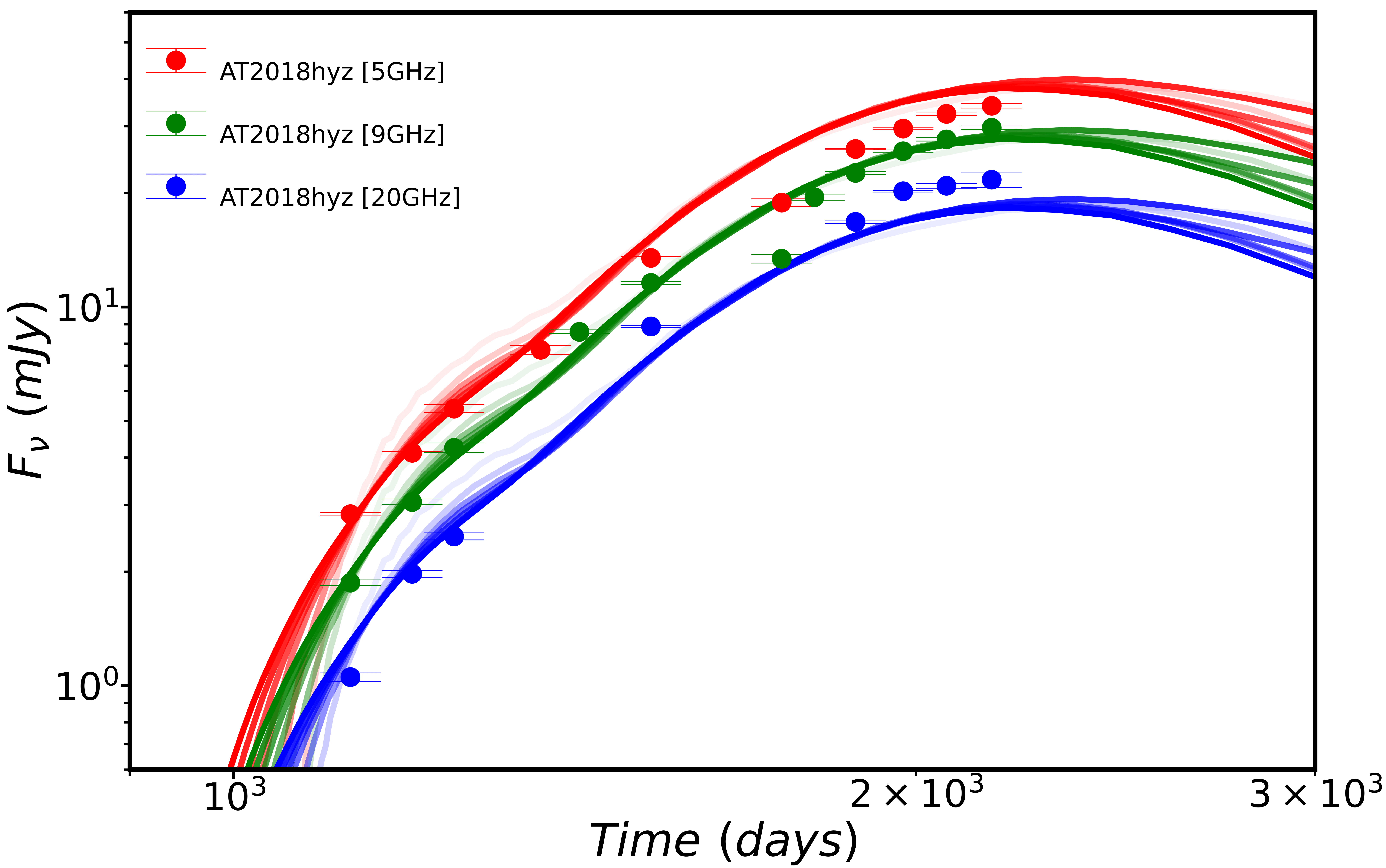}
\caption{\footnotesize{Simulated light curves and data comparison for three different radio bands (5 GHz in red, 9 GHz in green, and 20 GHz in blue). For each frequency, light curves for an off-axis observer ranging from $\theta_{obs} \in [30^{\circ},40^{\circ},50^{\circ},60^{\circ},70^{\circ},80^{\circ},90^{\circ}]$ with corresponding parameters (as mentioned in table \ref{tab:Thobs_Eiso_nism}) are shown with increasing transparency.}}
\label{fig:lc_fit_multi_Hz}
\end{figure}

\begin{table*}[!t]
\centering
\small
%\scriptsize
%\setlength{\tabcolsep}{2.5pt}

%\begin{adjustbox}{max width=\linewidth}
\begin{tabular}{c|ccc|ccc|ccc|ccc|ccc|ccc}
\hline
\multirow{3}{*}{$\theta_{\rm{obs}}$}
& \multicolumn{6}{c|}{3000 days (Dec 31, 2026)}
& \multicolumn{6}{c|}{3300 days (Oct 27, 2027)}
& \multicolumn{6}{c}{3600 days (Aug 22, 2028)} \\
\cline{2-19}

& \multicolumn{3}{c}{1 mJy}
& \multicolumn{3}{c|}{5 mJy}
& \multicolumn{3}{c}{1 mJy}
& \multicolumn{3}{c|}{5 mJy}
& \multicolumn{3}{c}{1 mJy}
& \multicolumn{3}{c}{5 mJy} \\
\cline{2-19}

& $X_c$ & $\Delta X$ & $\Delta Y$
& $X_c$ & $\Delta X$ & $\Delta Y$
& $X_c$ & $\Delta X$ & $\Delta Y$
& $X_c$ & $\Delta X$ & $\Delta Y$
& $X_c$ & $\Delta X$ & $\Delta Y$
& $X_c$ & $\Delta X$ & $\Delta Y$ \\
\hline

$30^{\circ}$
&5.70 & 0.23 & 0.52 & 5.60 & 0.64 & 1.69
&6.16 & 0.32 & 0.62 & 6.10 & 0.74 & 1.99
&6.69 & 0.36 & 1.17 & 6.54 & 1.02 & 2.43 \\

$40^{\circ}$
&4.08 & 0.18 & 0.48 & 4.02 & 0.45 & 1.16
&4.44 & 0.21 & 0.68 & 4.32 & 0.65 & 1.51
&4.71 & 0.27 & 0.86 & 4.55 & 0.93 & 2.07 \\

$50^{\circ}$
&3.16 & 0.14 & 0.39 & 3.10 & 0.41 & 0.94
&3.40 & 0.19 & 0.52 & 3.28 & 0.59 & 1.36
&3.60 & 0.26 & 0.71 & 3.39 & 0.91 & 1.89 \\

$60^{\circ}$
&2.57 & 0.12 & 0.32 & 2.51 & 0.36 & 0.78
&2.75 & 0.17 & 0.45 & 2.64 & 0.53 & 1.15
&2.89 & 0.23 & 0.61 & 2.74 & 0.68 & 1.42 \\

$70^{\circ}$
&2.08 & 0.12 & 0.31 & 2.04 & 0.29 & 0.63
&-1.56 & 0.17 & 0.31 & -1.52 & 0.34 & 0.59
&-1.72 & 0.17 & 0.33 & -1.66 & 0.44 & 0.76 \\

$80^{\circ}$
&-1.47 & 0.11 & 0.22 & -1.44 & 0.29 & 0.51
&-1.60 & 0.13 & 0.28 & -1.55 & 0.37 & 0.67
&-1.71 & 0.16 & 0.37 & -1.63 & 0.46 & 0.83 \\

$90^{\circ}$
&-1.45 & 0.13 & 0.29 & -1.40 & 0.31 & 0.62
&-1.53 & 0.16 & 0.42 & -1.46 & 0.42 & 0.75
&-1.59 & 0.20 & 0.46 & -1.47 & 0.57 & 1.09 \\

\hline
\end{tabular}
%\end{adjustbox}
\caption{\footnotesize{The size projection values for images shown in Fig. \ref{fig:sky_projection_firefly}. Sizes for contours enclosing 1 mJy and 5 mJy integrated flux are reported here. $X_c$, $\Delta X$, and $\Delta Y$ measure the centroid location, spread along the direction of apparent jet motion, and perpendicular to it in milliarcseconds, respectively. We report these values in intervals of around a year.}}
\label{tab:Sky_Projection_Data}
\end{table*}

\begin{figure*}
\centering
\includegraphics[width=1.0\textwidth]{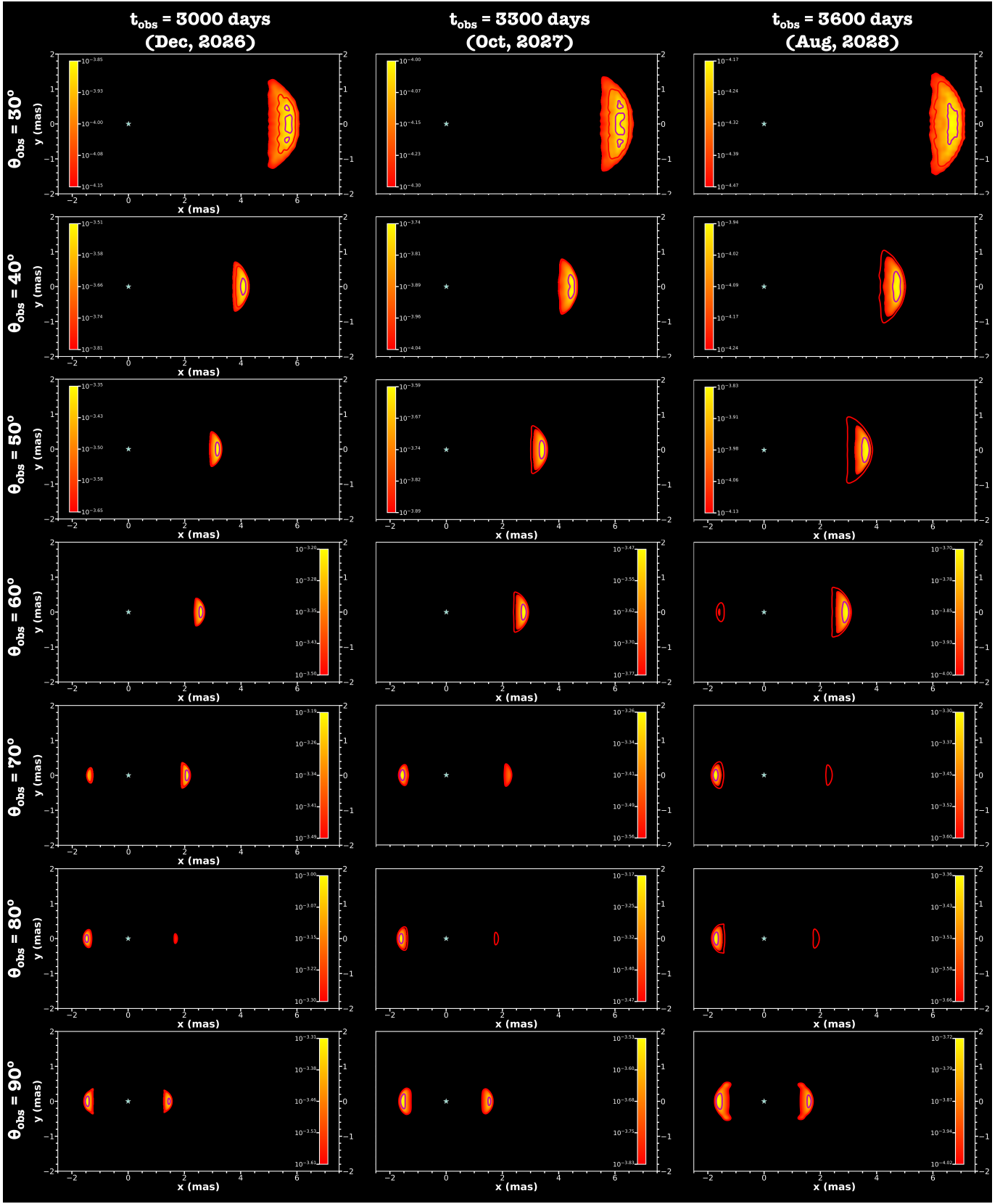}
\caption{\footnotesize{Sky projection images for different observer angles and times (since the optical disruption) for parameters listed in Table \ref{tab:Thobs_Eiso_nism}. These images are generated for $\nu_{\rm{obs}} = 9$ GHz and show contours enclosing 1 mJy and 5 mJy fluxes. The coordinates are given in milliarcseconds, with the source at z = 0.0457.}}
\label{fig:sky_projection_firefly}
\end{figure*}

We then generate sky projection images of the jet-ISM interaction flux as observed by an off-axis observer. The \textit{'Sky Projection'} feature of the \textit{Firefly} code, with the parameters reported above, was used for the sky images. Fig. \ref{fig:sky_projection_firefly} shows the 9 GHz flux maps for off-axis observers from Table \ref{tab:Thobs_Eiso_nism}, for the region bounded within the maximum flux and half of that value. We present sky projections at the end of this year, which is 3000 days since the disruption, and two more projections 300 days apart. These centroid images and motion can be traced by future VLBI observations. For direct comparison, we report the centroid size in terms of contours enclosing flux levels, as measured by VLBI. While VLBI at its highest sensitivity can resolve structures of 5 $\mu$Jy, it encapsulates a very small region for AT2018hyz. We report our sky projections as contours of structures enclosing 1 mJy and 5 mJy integrated flux. These structures are marked by red contours in Fig. \ref{fig:sky_projection_firefly}. We then calculate the flux-weighted centroid location ($X_c$) for the contours (Eq. \ref{Eq:Centroid_location}), their spread along the direction of jet motion ($\Delta X$), and perpendicular to it ($\Delta Y$). These values are reported in Table \ref{tab:Sky_Projection_Data}.

\section{Discussion} \label{sec:Discussion}
An observer viewing a jet off-axis naturally sees an earlier rise in light curves, followed by peak emission when the jet core becomes visible (that is, $\theta_{\rm{obs}} \sim 1/\Gamma_j$), and followed by a steep decline as the jet decelerates and spreads. While this can explain a delay in observing the light curve, off-axis viewing effects do not seem to account for the delayed radio brightening in AT2018hyz on their own. Our results indicate that the jet associated with this event may have been launched much later than the original tidal disruption of the star. This delay can arise from an initially misaligned TDE disk, because a precessing jet can remain confined by the disk wind until the inner accretion flow aligns with the black hole spin axis \citep{McKinney+2013, Teboul+2023}, or because of the fact that the returning debris does not necessarily form a compact accretion disk immediately and the circularization may take many orbital periods \citep{Piran+2015, Shiokawa+2015}. We conclude that these two effects, delayed jet launch and an off-axis viewing angle, together can explain the delayed radio rise in AT2018hyz.

\begin{figure}
\centering
\includegraphics[width=\linewidth]{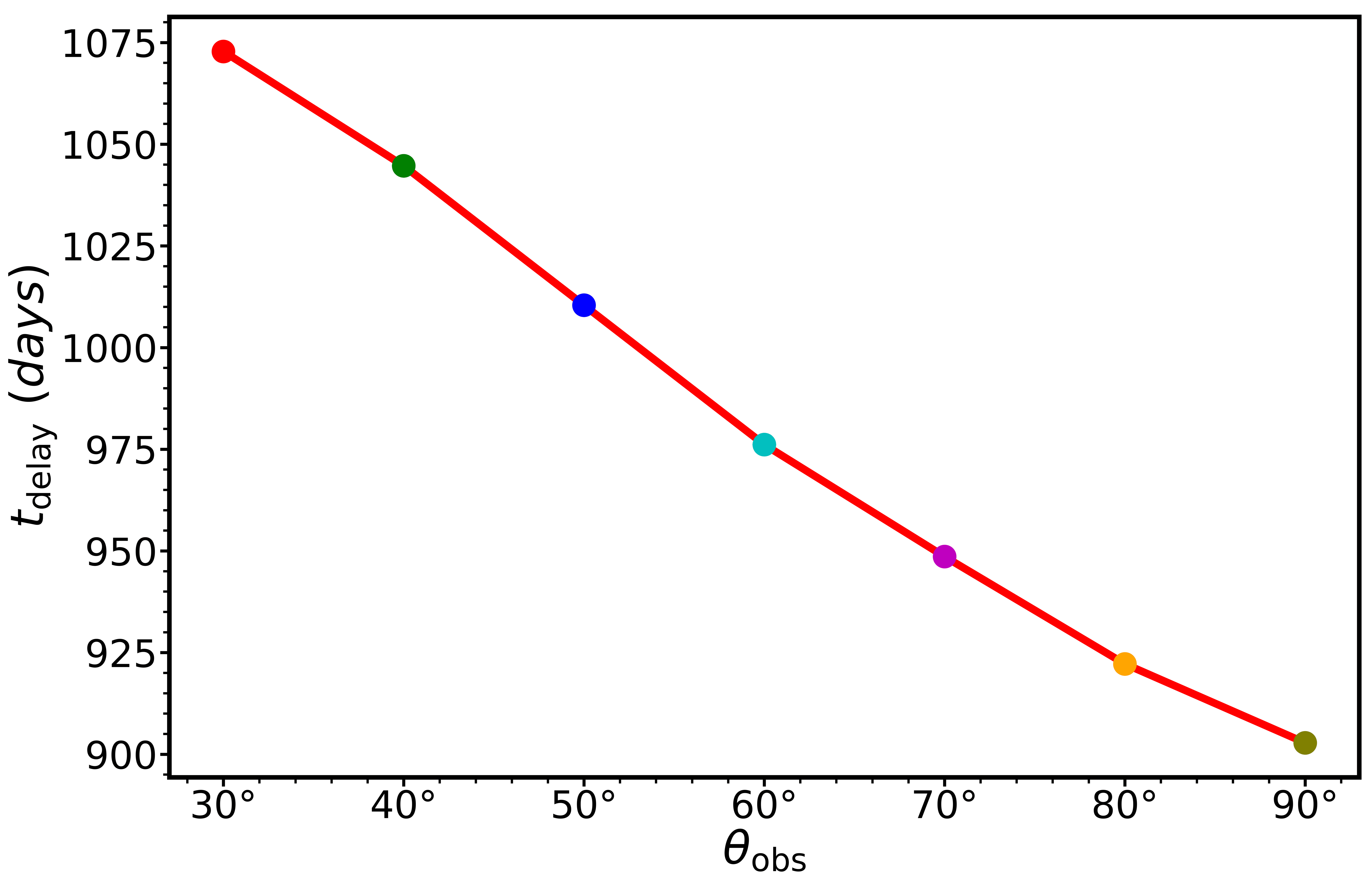}
\caption{\footnotesize{Correlation between observer angle and jet launch delay time required to explain the late radio afterglow emissions from AT2018hyz. A shorter delay in launching the jet would require the observer to be further off-axis. While such a correlation is expected to be universal, this particular curve results from MCMC done uniquely for observations of AT2018hyz.}}
\label{fig:thobs_tdelay}
\end{figure}

We find that the observer angle and the jet launch delay time (time since the tidal disruption) are inversely correlated, as shown in Fig \ref{fig:thobs_tdelay}. This implies that if the jet launched from AT2018hyz is observed far off-axis (around $\theta_{\rm{obs}} \approx 80^{\circ}$) would imply a shorter delay between the disruption and launching of the jet, compared to if the jet was observed closer to the axis (say, $\theta_{\rm{obs}} \approx 40^{\circ}$). Such a correlation between the observer angle and jet launch delay time in TDEs is expected to be universal. However, because the delay time is estimated from observed data for AT2018hyz, the curve shown in Fig. \ref{fig:thobs_tdelay} is specific to this TDE. In addition, we found that a lower isotropic equivalent energy and a denser ISM, along with a shorter delay time, are needed to explain the observations for a further off-axis jet. This is consistent with the energy-ISM density degeneracy in the afterglow theory. The $E_{\rm{iso}}$ and $n_{\rm{ISM}}$ range obtained in our study ranges from $\sim 2\times 10^{54}$ ergs, and $0.03 \rm{ cm}^{-3}$ respectively for a $30^{\circ}$ off-axis observer, to $\sim 3\times 10^{53}$, and $20 \rm{ cm}^{-3}$ respectively for an observer $90^{\circ}$ off-axis. These values are an order of magnitude higher than reported in \cite{Cendes+2026}, and lower than values obtained in \cite{Matsumoto+2025}, for the off-axis structured jet afterglow model without any delay in jet launching.

This gives us a set of possible $\theta_{\rm{obs}}$ and associated delay timescales (along with unique explosion energy and ISM density) to explain the observations in AT2018hyz. Light curves corresponding to these observer angles and associated parameters are not distinguishable from each other until $\sim 4\times 10^3$ days. However, since the jet spreads as it evolves, it exhibits different spreads in both directions (along the jet's motion and perpendicular to it) at different viewing angles. This is because of the changing equal arrival time surfaces. An observer closer to the jet core would see a larger fraction of the jet, while an observer further off-axis would see a smaller projection. For instance, in one extreme of $\theta_{\rm{obs}} = 0^{\circ}$ would see a circular disk slowing expanding with time as the jet spreads, while on the other extreme $\theta_{\rm{obs}} = 90^{\circ}$ would see a crescent shape for both the forward and counter jet equally bright and of the same size. Hence, sky projections can be used to break the observer-angle degeneracy in this model. We have presented the projected centroid size and spread for different observer angles, including the effects of jet spread. The further off-axis an observer is, the closer the sky image appears to the disruption center, with smaller spread and slower apparent motion. It can also be seen that further off-axis we go, the earlier we see the counter-jet emissions, as expected. Also, the centroid spread decreases in both $\Delta X$ and $\Delta Y$, thus projecting smaller and smaller images in the sky. Fig. \ref{fig:sky_projection_firefly} shows these projections.  We find that the smallest off-axis observer ($30^{\circ}$) sees the sky image furthest from the disruption location (5.6 mas at the 5 mJy contour) and also has the largest spread (0.64 mas). It also has the fastest superluminal apparent velocity, slightly decelerating from $\sim 3c$ to $\sim 2.5c$ over a span of 2500 days post-disruption (late 2024) to almost 4000 days (around one decade) post-explosion. Additionally, around $60^{\circ}$ off-axis, and after around 3600 days post disruption (that is, $\sim 2600$ days since jet launch), the counter jet becomes visible to the observer. This can be seen as a second bright region on the left of the disruption center. For observers beyond this angle, the counter jet appears earlier and becomes brighter than the forward jet over time. When accounting for jet spread and off-axis observers, the jet need not become completely spherical for the counter-jet to be visible (\cite{Dastidar+2024}). For an observer viewing the jet from the equatorial plane ($90^{\circ}$ off-axis), the forward and counter jets are indistinguishable, as expected, and we see two lobes of equal spread moving away from each other at the same apparent velocity.

An important caveat here is that the sky projections shown for different observer angles have different $E_{\rm{iso}}$, $n_{\rm{ISM}}$, and $t_{\rm{delay}}$ values. Hence, they cannot be compared directly. It may seem that the counter-jet dominates more at $\theta_{\rm{obs}} = 80^{\circ}$ than $\theta_{\rm{obs}} = 90^{\circ}$, which is counter intuitive. But it is due to the fact that the model for $80^{\circ}$ off-axis is also associated with a shorter delay time, along with higher $E_{\rm{iso}}$ and lower $n_{\rm{ISM}}$ than the $90^{\circ}$ off-axis model. However, given enough time, the counter jet would dominate emissions for all the observer angles (Fig. \ref{fig:lc_wo_w_delay}) but at different time scales, before they all emit fluxes below detection limits. Similarly, further-off-axis observers will report a slower centroid velocity and can be used to break the degeneracy. While there are currently no resolved sky projections reported for AT2018hyz, future VLBI observations can be used to narrow down the observer angle and, along with other parameters, the jet launch delay time.

Given the current upward trend in the light curve, it may not peak soon and will continue to rise; the simulated afterglow light curve can be adjusted by increasing $E_{\rm{iso}}$ and decreasing the ISM density to fit the data (for a given observer angle). Since the peak emission in the light curve also corresponds to the jet core being completely open to the observer, it would affect the sky projection. If the light curve continues to rise, it implies that the jet we observe is still well collimated. This, in turn, would imply that the equal arrival time surface, which effectively determines the size of the sky projection, would be smaller than we predict in this work (which predicts the peak should be soon). So, in effect, if the light curve continues to rise, the projection size we predict may be an upper limit based on current observations.

\section{Conclusion}\label{sec:conclusion}

In this paper, we investigated the late radio brightening observed from the tidal disruption event AT2018hyz. We explored the hypothesis that such late radio emission can originate from the afterglow of a jet-ISM interaction, with the delay attributable to off-axis observers and to an intrinsic delay in jet launch post the tidal disruption. We carried out a relativistic hydrodynamical jet evolution for a boosted fireball model for our jet. This also included accurately capturing the jet's lateral spread and its effects on the observations, especially the sky projection images. We post-process our hydrodynamical simulations with an afterglow code (\textit{Firefly}) for different observer angles at radio frequencies. Finally, we carried out MCMC analyses of the light curves for each observer angle to estimate parameters that best fit the observations. The key difference from other models for AT2018hyz is that, in our MCMC analysis, we also explore the parameter space for a possible delay between the disruption and the jet launch. Except for the parameters $\theta_{\rm{obs}}$, $E_{\rm{iso}}$, $n_{\rm{ISM}}$, and $t_{\rm{delay}}$ we use the values reported in previous literature \citep{Matsumoto+2025, Cendes+2026}.

We find that a jet launched around 900 to 1000 days after the tidal disruption was detected optically, viewed $90^{\circ}$ to $30^{\circ}$ off-axis, can successfully explain the radio observations. The further off-axis an observer is, the shorter the delay in launching the jet would correspond. In addition, different observer angles correspond to different isotropic-equivalent energy and ISM density. Hence, for each $\theta_{\rm{obs}}$ we find a unique set of $E_{\rm{iso}}$, $n_{\rm{ISM}}$, and $t_{\rm{delay}}$. This provides us with a range of models that produce degenerate light curves (until $\sim 4\times 10^3$ days), all of which satisfactorily match the observed fluxes.

This degeneracy among different observer angles can be broken by examining the jet's sky image. We predict the jet centroid spread and apparent motion for each observer angle for the corresponding parameters. We have presented the centroid sky images starting in Dec 2026, and for every 300 days since for two more epochs. These images can be directly observed with VLBI at radio frequencies. By comparing the observed sky images with these predictions, one should be able to narrow down the observer's angle and, hence, the delay between the disruption and the jet launch. The delay time is a significant probe into the accretion physics for tidal disruptions.

%Tidal disruption events are now being detected more frequently. Unlike traditional Gamma Ray Burst-associated jets, tidal disruptions accompanied by jets typically have optical precursors. Long term multi-band observations for such events can hence give further insights into the jet formation and launching mechanisms. Although the theory of jets in tidal disruption events is still in its early stages, we believe that much remains to be discovered and settled with more observations and theoretical models. This work has been one such attempt to correlate the late radio brightening in TDEs with an off-axis jet launched much later than the optically observed disruption. 

\begin{acknowledgments}

We acknowledge Yvette Cendes, Raffaella Margutti and Ryan Chornock for their helpful comments and insights in this work. The Grammarly AI tool has been used for spelling and grammar check in this manuscript. Hydrodynamical calculations were carried out on the Petunia cluster at Purdue University.  The development of the code \textit{Firefly} was supported by NASA under grant No. 80NSSC22K1615.
\end{acknowledgments}

\typeout{}
\bibliography{rana}

@article{Rees+1988b,
    author = "Rees, Martin J.",
    title = "{Tidal disruption of stars by black holes of 10 to the 6th-10 to the 8th solar masses in nearby galaxies}",
    doi = "10.1038/333523a0",
    journal = "Nature",
    volume = "333",
    pages = "523--528",
    year = "1988"
}

@ARTICLE{Komosa+2015,
       author = {{Komossa}, S.},
        title = "{Tidal disruption of stars by supermassive black holes: Status of observations}",
      journal = {Journal of High Energy Astrophysics},
         year = 2015,
        month = sep,
       volume = {7},
        pages = {148-157},
          doi = {10.1016/j.jheap.2015.04.006},
archivePrefix = {arXiv},
       eprint = {1505.01093},
 primaryClass = {astro-ph.HE},
       adsurl = {https://ui.adsabs.harvard.edu/abs/2015JHEAp...7..148K}
}

@ARTICLE{Gezari+2021,
       author = {{Gezari}, Suvi},
        title = "{Tidal Disruption Events}",
      journal = {\araa},
         year = 2021,
        month = sep,
       volume = {59},
        pages = {21-58},
          doi = {10.1146/annurev-astro-111720-030029},
archivePrefix = {arXiv},
       eprint = {2104.14580},
 primaryClass = {astro-ph.HE},
       adsurl = {https://ui.adsabs.harvard.edu/abs/2021ARA&A..59...21G}
}

@ARTICLE{Halpern+2004,
       author = {{Halpern}, J.~P. and {Gezari}, S. and {Komossa}, S.},
        title = "{Follow-Up Chandra Observations of Three Candidate Tidal Disruption Events}",
      journal = {\apj},
         year = 2004,
        month = apr,
       volume = {604},
       number = {2},
        pages = {572-578},
          doi = {10.1086/381937},
archivePrefix = {arXiv},
       eprint = {astro-ph/0402497},
 primaryClass = {astro-ph},
       adsurl = {https://ui.adsabs.harvard.edu/abs/2004ApJ...604..572H}
}

@article{Ladato+2011,
    author = {Lodato, Giuseppe and Rossi, Elena M.},
    title = {Multiband light curves of tidal disruption events},
    journal = {Monthly Notices of the Royal Astronomical Society},
    volume = {410},
    number = {1},
    pages = {359-367},
    year = {2011},
    month = {01},
    issn = {0035-8711},
    doi = {10.1111/j.1365-2966.2010.17448.x},
    url = {https://doi.org/10.1111/j.1365-2966.2010.17448.x},
    eprint = {https://academic.oup.com/mnras/article-pdf/410/1/359/18446097/mnras0410-0359.pdf},
}

@ARTICLE{Guillochon+2013,
       author = {{Guillochon}, James and {Ramirez-Ruiz}, Enrico},
        title = "{Hydrodynamical Simulations to Determine the Feeding Rate of Black Holes by the Tidal Disruption of Stars: The Importance of the Impact Parameter and Stellar Structure}",
      journal = {\apj},
         year = 2013,
        month = apr,
       volume = {767},
       number = {1},
          eid = {25},
        pages = {25},
          doi = {10.1088/0004-637X/767/1/25},
archivePrefix = {arXiv},
       eprint = {1206.2350},
 primaryClass = {astro-ph.HE},
       adsurl = {https://ui.adsabs.harvard.edu/abs/2013ApJ...767...25G}
}

@article{Law-Smith+2020,
doi = {10.3847/1538-4357/abc489},
url = {https://doi.org/10.3847/1538-4357/abc489},
year = {2020},
month = {dec},
publisher = {The American Astronomical Society},
volume = {905},
number = {2},
pages = {141},
author = {Law-Smith, Jamie A. P. and Coulter, David A. and Guillochon, James and Mockler, Brenna and Ramirez-Ruiz, Enrico},
title = {Stellar Tidal Disruption Events with Abundances and Realistic Structures (STARS): Library of Fallback Rates},
journal = {The Astrophysical Journal}
}

@ARTICLE{Ryu+2020,
       author = {{Ryu}, Taeho and {Krolik}, Julian and {Piran}, Tsvi and {Noble}, Scott C.},
        title = "{Tidal Disruptions of Main-sequence Stars. II. Simulation Methodology and Stellar Mass Dependence of the Character of Full Tidal Disruptions}",
      journal = {\apj},
         year = 2020,
        month = dec,
       volume = {904},
       number = {2},
          eid = {99},
        pages = {99},
          doi = {10.3847/1538-4357/abb3cd},
archivePrefix = {arXiv},
       eprint = {2001.03502},
 primaryClass = {astro-ph.HE},
       adsurl = {https://ui.adsabs.harvard.edu/abs/2020ApJ...904...99R}
}

@ARTICLE{Levan+2011,
       author = {{Levan}, A.~J. and {Tanvir}, N.~R. and {Cenko}, S.~B. and {Perley}, D.~A. and {Wiersema}, K. and {Bloom}, J.~S. and {Fruchter}, A.~S. and {de Ugarte Postigo}, A. and {O'Brien}, P.~T. and {Butler}, N. and {van der Horst}, A.~J. and {Leloudas}, G. and {Morgan}, A.~N. and {Misra}, K. and {Bower}, G.~C. and {Farihi}, J. and {Tunnicliffe}, R.~L. and {Modjaz}, M. and {Silverman}, J.~M. and {Hjorth}, J. and {Th{\"o}ne}, C. and {Cucchiara}, A. and {Cer{\'o}n}, J.~M. Castro and {Castro-Tirado}, A.~J. and {Arnold}, J.~A. and {Bremer}, M. and {Brodie}, J.~P. and {Carroll}, T. and {Cooper}, M.~C. and {Curran}, P.~A. and {Cutri}, R.~M. and {Ehle}, J. and {Forbes}, D. and {Fynbo}, J. and {Gorosabel}, J. and {Graham}, J. and {Hoffman}, D.~I. and {Guziy}, S. and {Jakobsson}, P. and {Kamble}, A. and {Kerr}, T. and {Kasliwal}, M.~M. and {Kouveliotou}, C. and {Kocevski}, D. and {Law}, N.~M. and {Nugent}, P.~E. and {Ofek}, E.~O. and {Poznanski}, D. and {Quimby}, R.~M. and {Rol}, E. and {Romanowsky}, A.~J. and {S{\'a}nchez-Ram{\'\i}rez}, R. and {Schulze}, S. and {Singh}, N. and {van Spaandonk}, L. and {Starling}, R.~L.~C. and {Strom}, R.~G. and {Tello}, J.~C. and {Vaduvescu}, O. and {Wheatley}, P.~J. and {Wijers}, R.~A.~M.~J. and {Winters}, J.~M. and {Xu}, D.},
        title = "{An Extremely Luminous Panchromatic Outburst from the Nucleus of a Distant Galaxy}",
      journal = {Science},
         year = 2011,
        month = jul,
       volume = {333},
       number = {6039},
        pages = {199},
          doi = {10.1126/science.1207143},
archivePrefix = {arXiv},
       eprint = {1104.3356},
 primaryClass = {astro-ph.HE},
       adsurl = {https://ui.adsabs.harvard.edu/abs/2011Sci...333..199L}
}

@ARTICLE{Zauderer+2011,
       author = {{Zauderer}, B.~A. and {Berger}, E. and {Soderberg}, A.~M. and {Loeb}, A. and {Narayan}, R. and {Frail}, D.~A. and {Petitpas}, G.~R. and {Brunthaler}, A. and {Chornock}, R. and {Carpenter}, J.~M. and {Pooley}, G.~G. and {Mooley}, K. and {Kulkarni}, S.~R. and {Margutti}, R. and {Fox}, D.~B. and {Nakar}, E. and {Patel}, N.~A. and {Volgenau}, N.~H. and {Culverhouse}, T.~L. and {Bietenholz}, M.~F. and {Rupen}, M.~P. and {Max-Moerbeck}, W. and {Readhead}, A.~C.~S. and {Richards}, J. and {Shepherd}, M. and {Storm}, S. and {Hull}, C.~L.~H.},
        title = "{Birth of a relativistic outflow in the unusual {\ensuremath{\gamma}}-ray transient Swift J164449.3+573451}",
      journal = {\nat},
         year = 2011,
        month = aug,
       volume = {476},
       number = {7361},
        pages = {425-428},
          doi = {10.1038/nature10366},
archivePrefix = {arXiv},
       eprint = {1106.3568},
 primaryClass = {astro-ph.HE},
       adsurl = {https://ui.adsabs.harvard.edu/abs/2011Natur.476..425Z}
}

@ARTICLE{Burrows+2011,
       author = {{Burrows}, D.~N. and {Kennea}, J.~A. and {Ghisellini}, G. and {Mangano}, V. and {Zhang}, B. and {Page}, K.~L. and {Eracleous}, M. and {Romano}, P. and {Sakamoto}, T. and {Falcone}, A.~D. and {Osborne}, J.~P. and {Campana}, S. and {Beardmore}, A.~P. and {Breeveld}, A.~A. and {Chester}, M.~M. and {Corbet}, R. and {Covino}, S. and {Cummings}, J.~R. and {D'Avanzo}, P. and {D'Elia}, V. and {Esposito}, P. and {Evans}, P.~A. and {Fugazza}, D. and {Gelbord}, J.~M. and {Hiroi}, K. and {Holland}, S.~T. and {Huang}, K.~Y. and {Im}, M. and {Israel}, G. and {Jeon}, Y. and {Jeon}, Y.-B. and {Jun}, H.~D. and {Kawai}, N. and {Kim}, J.~H. and {Krimm}, H.~A. and {Marshall}, F.~E. and {P. M{\'e}sz{\'a}ros} and {Negoro}, H. and {Omodei}, N. and {Park}, W.-K. and {Perkins}, J.~S. and {Sugizaki}, M. and {Sung}, H.-I. and {Tagliaferri}, G. and {Troja}, E. and {Ueda}, Y. and {Urata}, Y. and {Usui}, R. and {Antonelli}, L.~A. and {Barthelmy}, S.~D. and {Cusumano}, G. and {Giommi}, P. and {Melandri}, A. and {Perri}, M. and {Racusin}, J.~L. and {Sbarufatti}, B. and {Siegel}, M.~H. and {Gehrels}, N.},
        title = "{Relativistic jet activity from the tidal disruption of a star by a massive black hole}",
      journal = {\nat},
         year = 2011,
        month = aug,
       volume = {476},
       number = {7361},
        pages = {421-424},
          doi = {10.1038/nature10374},
archivePrefix = {arXiv},
       eprint = {1104.4787},
 primaryClass = {astro-ph.HE},
       adsurl = {https://ui.adsabs.harvard.edu/abs/2011Natur.476..421B}
}

@ARTICLE{Giannios+2011,
       author = {{Giannios}, Dimitrios and {Metzger}, Brian D.},
        title = "{Radio transients from stellar tidal disruption by massive black holes}",
      journal = {\mnras},
         year = 2011,
        month = sep,
       volume = {416},
       number = {3},
        pages = {2102-2107},
          doi = {10.1111/j.1365-2966.2011.19188.x},
archivePrefix = {arXiv},
       eprint = {1102.1429},
 primaryClass = {astro-ph.HE},
       adsurl = {https://ui.adsabs.harvard.edu/abs/2011MNRAS.416.2102G}
}

@ARTICLE{Bloom+2011,
       author = {{Bloom}, Joshua S. and {Giannios}, Dimitrios and {Metzger}, Brian D. and {Cenko}, S. Bradley and {Perley}, Daniel A. and {Butler}, Nathaniel R. and {Tanvir}, Nial R. and {Levan}, Andrew J. and {O'Brien}, Paul T. and {Strubbe}, Linda E. and {De Colle}, Fabio and {Ramirez-Ruiz}, Enrico and {Lee}, William H. and {Nayakshin}, Sergei and {Quataert}, Eliot and {King}, Andrew R. and {Cucchiara}, Antonino and {Guillochon}, James and {Bower}, Geoffrey C. and {Fruchter}, Andrew S. and {Morgan}, Adam N. and {van der Horst}, Alexander J.},
        title = "{A Possible Relativistic Jetted Outburst from a Massive Black Hole Fed by a Tidally Disrupted Star}",
      journal = {Science},
         year = 2011,
        month = jul,
       volume = {333},
       number = {6039},
        pages = {203},
          doi = {10.1126/science.1207150},
archivePrefix = {arXiv},
       eprint = {1104.3257},
 primaryClass = {astro-ph.HE},
       adsurl = {https://ui.adsabs.harvard.edu/abs/2011Sci...333..203B}
}

@article{DeColle+2012,
doi = {10.1088/0004-637X/760/2/103},
url = {https://doi.org/10.1088/0004-637X/760/2/103},
year = {2012},
month = {nov},
publisher = {The American Astronomical Society},
volume = {760},
number = {2},
pages = {103},
author = {De Colle, Fabio and Guillochon, James and Naiman, Jill and Ramirez-Ruiz, Enrico},
title = {THE DYNAMICS, APPEARANCE, AND DEMOGRAPHICS OF RELATIVISTIC JETS TRIGGERED BY TIDAL DISRUPTION OF STARS IN QUIESCENT SUPERMASSIVE BLACK HOLES},
journal = {The Astrophysical Journal}
}

@article{Bower+2013,
doi = {10.1088/0004-637X/763/2/84},
url = {https://doi.org/10.1088/0004-637X/763/2/84},
year = {2013},
month = {jan},
publisher = {The American Astronomical Society},
volume = {763},
number = {2},
pages = {84},
author = {Bower, Geoffrey C. and Metzger, Brian D. and Cenko, S. Bradley and Silverman, Jeffrey M. and Bloom, Joshua S.},
title = {LATE-TIME RADIO EMISSION FROM X-RAY-SELECTED TIDAL DISRUPTION EVENTS},
journal = {The Astrophysical Journal}
}

@article{Brown+2015,
    author = {Brown, G. C. and Levan, A. J. and Stanway, E. R. and Tanvir, N. R. and Cenko, S. B. and Berger, E. and Chornock, R. and Cucchiaria, A.},
    title = {Swift J1112.2‚àí8238: a candidate relativistic tidal disruption flare},
    journal = {Monthly Notices of the Royal Astronomical Society},
    volume = {452},
    number = {4},
    pages = {4297-4306},
    year = {2015},
    month = {10},
    issn = {0035-8711},
    doi = {10.1093/mnras/stv1520},
    url = {https://doi.org/10.1093/mnras/stv1520},
    eprint = {https://academic.oup.com/mnras/article-pdf/452/4/4297/18238481/stv1520.pdf},
}

@ARTICLE{Velzen+2013,
       author = {{van Velzen}, S. and {Frail}, D.~A. and {K{\"o}rding}, E. and {Falcke}, H.},
        title = "{Constraints on off-axis jets from stellar tidal disruption flares}",
      journal = {\aap},
         year = 2013,
        month = apr,
       volume = {552},
          eid = {A5},
        pages = {A5},
          doi = {10.1051/0004-6361/201220426},
archivePrefix = {arXiv},
       eprint = {1210.0022},
 primaryClass = {astro-ph.HE},
       adsurl = {https://ui.adsabs.harvard.edu/abs/2013A&A...552A...5V}
}

@article{Alexander+2016,
doi = {10.3847/2041-8205/819/2/L25},
url = {https://doi.org/10.3847/2041-8205/819/2/L25},
year = {2016},
month = {mar},
publisher = {The American Astronomical Society},
volume = {819},
number = {2},
pages = {L25},
author = {Alexander, K. D. and Berger, E. and Guillochon, J. and Zauderer, B. A. and Williams, P. K. G.},
title = {DISCOVERY OF AN OUTFLOW FROM RADIO OBSERVATIONS OF THE TIDAL DISRUPTION EVENT ASASSN-14li},
journal = {The Astrophysical Journal Letters}
}

@article{Cendes+2021,
doi = {10.3847/1538-4357/ac110a},
url = {https://doi.org/10.3847/1538-4357/ac110a},
year = {2021},
month = {oct},
publisher = {The American Astronomical Society},
volume = {919},
number = {2},
pages = {127},
author = {Cendes, Y. and Alexander, K. D. and Berger, E. and Eftekhari, T. and Williams, P. K. G. and Chornock, R.},
title = {Radio Observations of an Ordinary Outflow from the Tidal Disruption Event AT2019dsg},
journal = {The Astrophysical Journal}
}

@article{Stein+2020,
    author = "Stein, Robert and others",
    title = "{A tidal disruption event coincident with a high-energy neutrino}",
    eprint = "2005.05340",
    archivePrefix = "arXiv",
    primaryClass = "astro-ph.HE",
    doi = "10.1038/s41550-020-01295-8",
    journal = "Nature Astron.",
    volume = "5",
    number = "5",
    pages = "510--518",
    year = "2021"
}

@ARTICLE{Cendes+2024,
       author = {{Cendes}, Y. and {Berger}, E. and {Alexander}, K.~D. and {Chornock}, R. and {Margutti}, R. and {Metzger}, B. and {Wieringa}, M.~H. and {Bietenholz}, M.~F. and {Hajela}, A. and {Laskar}, T. and {Stroh}, M.~C. and {Terreran}, G.},
        title = "{Ubiquitous Late Radio Emission from Tidal Disruption Events}",
      journal = {\apj},
         year = 2024,
        month = aug,
       volume = {971},
       number = {2},
          eid = {185},
        pages = {185},
          doi = {10.3847/1538-4357/ad5541},
archivePrefix = {arXiv},
       eprint = {2308.13595},
 primaryClass = {astro-ph.HE},
       adsurl = {https://ui.adsabs.harvard.edu/abs/2024ApJ...971..185C}
}

@article{Mou+2022,
    author = {Mou, Guobin and Wang, Tinggui and Wang, Wei and Yang, Jingjing},
    title = {Radio emission from outflow‚Äìcloud interaction and its constraint on tidal disruption event outflow},
    journal = {Monthly Notices of the Royal Astronomical Society},
    volume = {510},
    number = {3},
    pages = {3650-3657},
    year = {2022},
    month = {03},
    issn = {0035-8711},
    doi = {10.1093/mnras/stab3742},
    url = {https://doi.org/10.1093/mnras/stab3742},
    eprint = {https://academic.oup.com/mnras/article-pdf/510/3/3650/42151820/stab3742.pdf},
}

@ARTICLE{Zhuang+2025,
       author = {{Zhuang}, Jialun and {Shen}, Rong-Feng and {Mou}, Guobin and {Lu}, Wenbin},
        title = "{Interaction of an Outflow with Surrounding Gaseous Clouds as the Origin of Late-time Radio Flares in Tidal Disruption Events}",
      journal = {\apj},
         year = 2025,
        month = feb,
       volume = {979},
       number = {2},
          eid = {109},
        pages = {109},
          doi = {10.3847/1538-4357/ad9b98},
archivePrefix = {arXiv},
       eprint = {2406.08012},
 primaryClass = {astro-ph.HE},
       adsurl = {https://ui.adsabs.harvard.edu/abs/2025ApJ...979..109Z}
}

@article{Matsumoto+2024,
doi = {10.3847/1538-4357/ad58ba},
url = {https://doi.org/10.3847/1538-4357/ad58ba},
year = {2024},
month = {aug},
publisher = {The American Astronomical Society},
volume = {971},
number = {1},
pages = {49},
author = {Matsumoto, Tatsuya and Piran, Tsvi},
title = {Late-time Radio Flares in Tidal Disruption Events},
journal = {The Astrophysical Journal}
}

@ARTICLE{Goodwin+2025,
       author = {{Goodwin}, A.~J. and {Mummery}, A. and {Laskar}, T. and {Alexander}, K.~D. and {Anderson}, G.~E. and {Bietenholz}, M. and {Bonnerot}, C. and {Christy}, C.~T. and {Golay}, W. and {Lu}, W. and {Margutti}, R. and {Miller-Jones}, J.~C.~A. and {Ramirez-Ruiz}, E. and {Saxton}, R. and {van Velzen}, S.},
        title = "{A Second Radio Flare from the Tidal Disruption Event AT2020vwl: A Delayed Outflow Ejection?}",
      journal = {\apj},
         year = 2025,
        month = mar,
       volume = {981},
       number = {2},
          eid = {122},
        pages = {122},
          doi = {10.3847/1538-4357/adb0b1},
archivePrefix = {arXiv},
       eprint = {2410.18665},
 primaryClass = {astro-ph.HE},
       adsurl = {https://ui.adsabs.harvard.edu/abs/2025ApJ...981..122G}
}

@article{Goodwin+2023,
    author = {Goodwin, A J and Alexander, K D and Miller-Jones, J C A and Bietenholz, M F and van¬†Velzen, S and Anderson, G E and Berger, E and Cendes, Y and Chornock, R and Coppejans, D L and Eftekhari, T and Gezari, S and Laskar, T and Ramirez-Ruiz, E and Saxton, R},
    title = {A radio-emitting outflow produced by the tidal disruption event AT2020vwl},
    journal = {Monthly Notices of the Royal Astronomical Society},
    volume = {522},
    number = {4},
    pages = {5084-5097},
    year = {2023},
    month = {07},
    issn = {0035-8711},
    doi = {10.1093/mnras/stad1258},
    url = {https://doi.org/10.1093/mnras/stad1258},
    eprint = {https://academic.oup.com/mnras/article-pdf/522/4/5084/50311114/stad1258.pdf},
}

@ARTICLE{Hajela+2025,
       author = {{Hajela}, A. and {Alexander}, K.~D. and {Margutti}, R. and {Chornock}, R. and {Bietenholz}, M. and {Christy}, C.~T. and {Stroh}, M. and {Terreran}, G. and {Saxton}, R. and {Komossa}, S. and {Bright}, J.~S. and {Ramirez-Ruiz}, E. and {Coppejans}, D.~L. and {Leung}, J.~K. and {Cendes}, Y. and {Wiston}, E. and {Laskar}, T. and {Horesh}, A. and {Schroeder}, G. and {A.~J.}, Nayana and {Wieringa}, M.~H. and {Velez}, N. and {Berger}, E. and {Blanchard}, P.~K. and {Eftekhari}, T. and {Gomez}, S. and {Nicholl}, M. and {Sears}, H. and {Zauderer}, B.~A.},
        title = "{Eight Years of Light from ASASSN-15oi: Toward Understanding the Late-time Evolution of TDEs}",
      journal = {\apj},
         year = 2025,
        month = apr,
       volume = {983},
       number = {1},
          eid = {29},
        pages = {29},
          doi = {10.3847/1538-4357/adb620},
archivePrefix = {arXiv},
       eprint = {2407.19019},
 primaryClass = {astro-ph.HE},
       adsurl = {https://ui.adsabs.harvard.edu/abs/2025ApJ...983...29H}
}

@article{Matsumoto+2023,
    author = {Matsumoto, Tatsuya and Piran, Tsvi},
    title = {Generalized equipartition method from an arbitrary viewing angle},
    journal = {Monthly Notices of the Royal Astronomical Society},
    volume = {522},
    number = {3},
    pages = {4565-4576},
    year = {2023},
    month = {07},
    issn = {0035-8711},
    doi = {10.1093/mnras/stad1269},
    url = {https://doi.org/10.1093/mnras/stad1269},
    eprint = {https://academic.oup.com/mnras/article-pdf/522/3/4565/50267991/stad1269.pdf},
}

@article{Sfaradi+2024,
    author = {Sfaradi, Itai and Beniamini, Paz and Horesh, Assaf and Piran, Tsvi and Bright, Joe and Rhodes, Lauren and Williams, David R A and Fender, Rob and Leung, James K and Murphy, Tara and Green, Dave A},
    title = {An off-axis relativistic jet seen in the long lasting delayed radio flare of the TDE AT 2018hyz},
    journal = {Monthly Notices of the Royal Astronomical Society},
    volume = {527},
    number = {3},
    pages = {7672-7680},
    year = {2024},
    month = {01},
    issn = {0035-8711},
    doi = {10.1093/mnras/stad3717},
    url = {https://doi.org/10.1093/mnras/stad3717},
    eprint = {https://academic.oup.com/mnras/article-pdf/527/3/7672/54357442/stad3717.pdf},
}

@article{Hayasaki+2021,
doi = {10.3847/1538-4357/ac18c2},
url = {https://doi.org/10.3847/1538-4357/ac18c2},
year = {2021},
month = {oct},
publisher = {The American Astronomical Society},
volume = {921},
number = {1},
pages = {20},
author = {Hayasaki, Kimitake and Jonker, Peter G.},
title = {On the Origin of Late-time X-Ray Flares in UV/optically Selected Tidal Disruption Events},
journal = {The Astrophysical Journal}
}

@ARTICLE{Horesh+2021,
       author = {{Horesh}, A. and {Cenko}, S.~B. and {Arcavi}, I.},
        title = "{Delayed radio flares from a tidal disruption event}",
      journal = {Nature Astronomy},
         year = 2021,
        month = may,
       volume = {5},
        pages = {491-497},
          doi = {10.1038/s41550-021-01300-8},
archivePrefix = {arXiv},
       eprint = {2102.11290},
 primaryClass = {astro-ph.HE},
       adsurl = {https://ui.adsabs.harvard.edu/abs/2021NatAs...5..491H}
}

@article{Cendes+2022,
doi = {10.3847/1538-4357/ac88d0},
url = {https://doi.org/10.3847/1538-4357/ac88d0},
year = {2022},
month = {oct},
publisher = {The American Astronomical Society},
volume = {938},
number = {1},
pages = {28},
author = {Cendes, Y. and Berger, E. and Alexander, K. D. and Gomez, S. and Hajela, A. and Chornock, R. and Laskar, T. and Margutti, R. and Metzger, B. and Bietenholz, M. F. and Brethauer, D. and Wieringa, M. H.},
title = {A Mildly Relativistic Outflow Launched Two Years after Disruption in Tidal Disruption Event AT2018hyz},
journal = {The Astrophysical Journal}
}

@article{Zhang+2024,
doi = {10.3847/2041-8213/ad1d61},
url = {https://doi.org/10.3847/2041-8213/ad1d61},
year = {2024},
month = {feb},
publisher = {The American Astronomical Society},
volume = {962},
number = {1},
pages = {L18},
author = {Zhang, Fabao and Shu, Xinwen and Yang, Lei and Sun, Luming and Zhang, Zhumao and Wang, Yibo and Mou, Guobin and Zhang, Xue-Guang and Zhou, Tianyao and Peng, Fangkun},
title = {Delayed and Fast-rising Radio Flares from an Optical and X-Ray-detected Tidal Disruption Event in the Center of a Dwarf Galaxy},
journal = {The Astrophysical Journal Letters}
}

@ARTICLE{Alexander+2026,
       author = {{Alexander}, Kate D. and {Margutti}, Raffaella and {Gomez}, Sebastian and {Stroh}, Michael and {Chornock}, Ryan and {Laskar}, Tanmoy and {Cendes}, Y. and {Berger}, Edo and {Eftekhari}, Tarraneh and {Franz}, Noah and {Hajela}, Aprajita and {Metzger}, B.~D. and {Terreran}, Giacomo and {Bietenholz}, Michael and {Christy}, Collin and {De Colle}, Fabio and {Komossa}, S. and {Nicholl}, Matt and {Ramirez-Ruiz}, Enrico and {Saxton}, Richard and {Schroeder}, Genevieve and {Williams}, Peter K.~G. and {Wu}, William},
        title = "{The Multiwavelength Context of Delayed Radio Emission in Tidal Disruption Events: Evidence for Accretion-driven Outflows}",
      journal = {\apj},
         year = 2026,
        month = mar,
       volume = {1000},
       number = {1},
          eid = {139},
        pages = {139},
          doi = {10.3847/1538-4357/ae40ab},
archivePrefix = {arXiv},
       eprint = {2506.12729},
 primaryClass = {astro-ph.HE},
       adsurl = {https://ui.adsabs.harvard.edu/abs/2026ApJ..1000..139A}
}

@article{Cendes+2026,
doi = {10.3847/1538-4357/ae286d},
url = {https://doi.org/10.3847/1538-4357/ae286d},
year = {2026},
month = {feb},
publisher = {The American Astronomical Society},
volume = {998},
number = {1},
pages = {111},
author = {Cendes, Yvette and Berger, Edo and Beniamini, Paz and Gill, Ramandeep and Matsumoto, Tatsuya and Alexander, Kate D. and Bietenholz, Michael F. and Hajela, Aprajita and Christy, Collin T. and Chornock, Ryan and Gomez, Sebastian and Gurwell, Mark A. and Keating, Garrett K. and Laskar, Tanmoy and Margutti, Raffaella and Rao, Ramprasad and Velez, Natalie and Wieringa, Mark H.},
title = {Continued Rapid Radio Brightening of the Tidal Disruption Event AT2018hyz},
journal = {The Astrophysical Journal}
}

@ARTICLE{Teboul+2023,
       author = {{Teboul}, Odelia and {Metzger}, Brian D.},
        title = "{A Unified Theory of Jetted Tidal Disruption Events: From Promptly Escaping Relativistic to Delayed Transrelativistic Jets}",
      journal = {\apjl},
         year = 2023,
        month = nov,
       volume = {957},
       number = {1},
          eid = {L9},
        pages = {L9},
          doi = {10.3847/2041-8213/ad0037},
archivePrefix = {arXiv},
       eprint = {2308.05161},
 primaryClass = {astro-ph.HE},
       adsurl = {https://ui.adsabs.harvard.edu/abs/2023ApJ...957L...9T}
}

@ARTICLE{Goley+2025,
       author = {{Golay}, Walter W. and {Berger}, Edo and {Cendes}, Yvette and {Masterson}, Megan and {Polisensky}, Emil and {Mutel}, Robert L. and {Blanchard}, Peter K. and {Kumar}, Harsh and {Margutti}, Raffaella and {Drout}, Maria and {Panagiotou}, Christos and {De}, Kishalay and {Kara}, Erin},
        title = "{Radio Emission from the Infrared Tidal Disruption Event WTP14adeqka: The First Directly Resolved Delayed Outflow from a TDE}",
      journal = {arXiv e-prints},
         year = 2025,
        month = aug,
          eid = {arXiv:2508.16756},
        pages = {arXiv:2508.16756},
          doi = {10.48550/arXiv.2508.16756},
archivePrefix = {arXiv},
       eprint = {2508.16756},
 primaryClass = {astro-ph.HE},
       adsurl = {https://ui.adsabs.harvard.edu/abs/2025arXiv250816756G}
}

@article{Christy+2024,
doi = {10.3847/1538-4357/ad675b},
url = {https://doi.org/10.3847/1538-4357/ad675b},
year = {2024},
month = {oct},
publisher = {The American Astronomical Society},
volume = {974},
number = {1},
pages = {18},
author = {Christy, Collin T. and Alexander, Kate D. and Margutti, Raffaella and Wieringa, Mark and Cendes, Yvette and Chornock, Ryan and Laskar, Tanmoy and Berger, Edo and Bietenholz, Michael and Coppejans, Deanne L. and De Colle, Fabio and Eftekhari, Tarraneh and Holoien, Thomas W.-S. and Matsumoto, Tatsuya and Miller-Jones, James C. A. and Ramirez-Ruiz, Enrico and Saxton, Richard and van Velzen, Sjoert},
title = {The Peculiar Radio Evolution of the Tidal Disruption Event ASASSN-19bt},
journal = {The Astrophysical Journal}
}

@ARTICLE{Matsumoto+2025,
       author = {{Matsumoto}, Tatsuya},
        title = "{VLBI Diagnostics of Off-axis Jets in Radio Flares of Tidal Disruption Events}",
      journal = {arXiv e-prints},
         year = 2025,
        month = dec,
          eid = {arXiv:2512.21669},
        pages = {arXiv:2512.21669},
          doi = {10.48550/arXiv.2512.21669},
archivePrefix = {arXiv},
       eprint = {2512.21669},
 primaryClass = {astro-ph.HE},
       adsurl = {https://ui.adsabs.harvard.edu/abs/2025arXiv251221669M}
}

@ARTICLE{Wu+2026,
       author = {{Wu}, Samantha C. and {Tsuna}, Daichi and {Mockler}, Brenna and {Piro}, Anthony L.},
        title = "{Delayed Radio Emission in Tidal Disruption Events from Collisions of Outflows Driven by Disk Instabilities}",
      journal = {\apj},
         year = 2026,
        month = feb,
       volume = {998},
       number = {2},
          eid = {199},
        pages = {199},
          doi = {10.3847/1538-4357/ae36a3},
archivePrefix = {arXiv},
       eprint = {2511.14008},
 primaryClass = {astro-ph.HE},
       adsurl = {https://ui.adsabs.harvard.edu/abs/2026ApJ...998..199W}
}

@ARTICLE{Sato+2024,
       author = {{Sato}, Yuri and {Murase}, Kohta and {Bhattacharya}, Mukul and {Carpio}, Jose A. and {Mukhopadhyay}, Mainak and {Zhang}, B. Theodore},
        title = "{Two-component off-axis jet model for radio flares of tidal disruption events}",
      journal = {\prd},
         year = 2024,
        month = sep,
       volume = {110},
       number = {6},
          eid = {L061307},
        pages = {L061307},
          doi = {10.1103/PhysRevD.110.L061307},
archivePrefix = {arXiv},
       eprint = {2404.13326},
 primaryClass = {astro-ph.HE},
       adsurl = {https://ui.adsabs.harvard.edu/abs/2024PhRvD.110f1307S}
}

@ARTICLE{Horesh+2018,
       author = {{Horesh}, A. and {Sfaradi}, I. and {Bright}, J. and {Williams}, D. and {Fender}, R. and {Titterington}, D. and {Green}, D. and {Perrott}, Y. and {Van Velzen}, S. and {Gezari}, S.},
        title = "{AMI-LA 15.5 GHz observation of the TDE candidate AT2018HCO}",
      journal = {The Astronomer's Telegram},
         year = 2018,
        month = dec,
       volume = {12271},
        pages = {1},
       adsurl = {https://ui.adsabs.harvard.edu/abs/2018ATel12271....1H}
}

@ARTICLE{Gomez+2020,
       author = {{Gomez}, Sebastian and {Nicholl}, Matt and {Short}, Philip and {Margutti}, Raffaella and {Alexander}, Kate D. and {Blanchard}, Peter K. and {Berger}, Edo and {Eftekhari}, Tarraneh and {Schulze}, Steve and {Anderson}, Joseph and {Arcavi}, Iair and {Chornock}, Ryan and {Cowperthwaite}, Philip S. and {Galbany}, Llu{\'\i}s and {Herzog}, Laura J. and {Hiramatsu}, Daichi and {Hosseinzadeh}, Griffin and {Laskar}, Tanmoy and {M{\"u}ller Bravo}, Tom{\'a}s E. and {Patton}, Locke and {Terreran}, Giacomo},
        title = "{The Tidal Disruption Event AT 2018hyz II: Light-curve modelling of a partially disrupted star}",
      journal = {\mnras},
         year = 2020,
        month = sep,
       volume = {497},
       number = {2},
        pages = {1925-1934},
          doi = {10.1093/mnras/staa2099},
archivePrefix = {arXiv},
       eprint = {2003.05469},
 primaryClass = {astro-ph.HE},
       adsurl = {https://ui.adsabs.harvard.edu/abs/2020MNRAS.497.1925G}
}

@article{Duffell+2013,
doi = {10.1088/2041-8205/776/1/L9},
url = {https://dx.doi.org/10.1088/2041-8205/776/1/L9},
year = {2013},
month = {sep},
publisher = {The American Astronomical Society},
volume = {776},
number = {1},
pages = {L9},
author = {Paul C. Duffell and Andrew I. MacFadyen},
title = {A “BOOSTED FIREBALL” MODEL FOR STRUCTURED RELATIVISTIC JETS},
journal = {The Astrophysical Journal Letters}
}

@ARTICLE{Duffell+2018,
doi = {10.3847/1538-4357/aae084},
url = {https://dx.doi.org/10.3847/1538-4357/aae084},
year = {2018},
month = {oct},
publisher = {The American Astronomical Society},
volume = {866},
number = {1},
pages = {3},
author = {Paul C. Duffell and Eliot Quataert and Daniel Kasen and Hannah Klion},
title = {Jet Dynamics in Compact Object Mergers: GW170817 Likely Had a Successful Jet},
journal = {The Astrophysical Journal},
}

@ARTICLE{Wu+2018,
       author = {{Wu}, Yiyang and {MacFadyen}, Andrew},
        title = "{Constraining the Outflow Structure of the Binary Neutron Star Merger Event GW170817/GRB170817A with a Markov Chain Monte Carlo Analysis}",
      journal = {\apj},
         year = 2018,
        month = dec,
       volume = {869},
       number = {1},
          eid = {55},
        pages = {55},
          doi = {10.3847/1538-4357/aae9de},
archivePrefix = {arXiv},
       eprint = {1809.06843},
 primaryClass = {astro-ph.HE},
       adsurl = {https://ui.adsabs.harvard.edu/abs/2018ApJ...869...55W}
}

@article{Dastidar+2024,
doi = {10.3847/1538-4357/ad86bf},
url = {https://dx.doi.org/10.3847/1538-4357/ad86bf},
year = {2024},
month = {nov},
publisher = {The American Astronomical Society},
volume = {976},
number = {2},
pages = {252},
author = {Ranadeep G. Dastidar and Paul C. Duffell},
title = {Could the Recent Rebrightening of the GW170817A Afterglow Be Caused by a Counterjet?},
journal = {The Astrophysical Journal}
}

@article{Schroeder+2025,
    author = "Schroeder, Genevieve and Ho, Anna Y. Q. and Dastidar, Ranadeep G. and Modjaz, Maryam and Corsi, Alessandra and Duffell, Paul C.",
    title = "{A Late-time Radio Search for Highly Off-axis Jets from PTF Broad-lined Ic Supernovae in GRB-like Host Galaxy Environments}",
    eprint = "2507.15928",
    archivePrefix = "arXiv",
    primaryClass = "astro-ph.HE",
    doi = "10.3847/1538-4357/ae129b",
    journal = "Astrophys. J.",
    volume = "995",
    number = "1",
    pages = "61",
    year = "2025"
}

@article{Wang+2024,
    author = "Wang, Hao and Dastidar, Ranadeep G. and Giannios, Dimitrios and Duffell, Paul C.",
    title = "{jetsimpy: A Highly Efficient Hydrodynamic Code for Gamma-Ray Burst Afterglow}",
    eprint = "2402.19359",
    archivePrefix = "arXiv",
    primaryClass = "astro-ph.HE",
    doi = "10.3847/1538-4365/ad4d9d",
    journal = "Astrophys. J. Suppl.",
    volume = "273",
    number = "1",
    pages = "17",
    year = "2024"
}

@article{Sari+1998,
doi = {10.1086/311269},
url = {https://dx.doi.org/10.1086/311269},
year = {1998},
month = {mar},
publisher = {},
volume = {497},
number = {1},
pages = {L17},
author = {Re'em Sari and Tsvi Piran and Ramesh Narayan},
title = {Spectra and Light Curves of Gamma-Ray Burst Afterglows},
journal = {The Astrophysical Journal}
}

@article{Granot+2002,
doi = {10.1086/338966},
url = {https://dx.doi.org/10.1086/338966},
year = {2002},
month = {apr},
publisher = {},
volume = {568},
number = {2},
pages = {820},
author = {Jonathan Granot and Re’em Sari},
title = {The Shape of Spectral Breaks in Gamma-Ray Burst Afterglows},
journal = {The Astrophysical Journal}
}

@ARTICLE{Eerten+2013,
       author = {{van Eerten}, Hendrik},
        title = "{Gamma-ray burst afterglow theory}",
      journal = {arXiv e-prints},
         year = 2013,
        month = sep,
          eid = {arXiv:1309.3869},
        pages = {arXiv:1309.3869},
          doi = {10.48550/arXiv.1309.3869},
archivePrefix = {arXiv},
       eprint = {1309.3869},
 primaryClass = {astro-ph.HE},
       adsurl = {https://ui.adsabs.harvard.edu/abs/2013arXiv1309.3869V}
}

@article{Eerten+2024,
    author = {{van Eerten}, Hendrik J. and {Ryan}, Geoffrey S.},
    title = {Scaling relations for gamma-ray burst afterglow light curves and centroid motion independent of jet structure and dynamics},
    journal = {Monthly Notices of the Royal Astronomical Society},
    volume = {530},
    number = {4},
    pages = {4094-4106},
    year = {2024},
    month = {06},
    issn = {0035-8711},
    doi = {10.1093/mnras/stae1128},
    url = {https://doi.org/10.1093/mnras/stae1128},
    eprint = {https://academic.oup.com/mnras/article-pdf/530/4/4094/57439677/stae1128.pdf},
}

@article{McKinney+2013,
author = {Jonathan C. McKinney  and Alexander Tchekhovskoy  and Roger D. Blandford },
title = {Alignment of Magnetized Accretion Disks and Relativistic Jets with Spinning Black Holes},
journal = {Science},
volume = {339},
number = {6115},
pages = {49-52},
year = {2013},
doi = {10.1126/science.1230811},
URL = {https://www.science.org/doi/abs/10.1126/science.1230811},
eprint = {https://www.science.org/doi/pdf/10.1126/science.1230811}}

@ARTICLE{Piran+2015,
       author = {{Piran}, Tsvi and {Svirski}, Gilad and {Krolik}, Julian and {Cheng}, Roseanne M. and {Shiokawa}, Hotaka},
        title = "{′Disk Formation Versus Disk Accretion{\textemdash}What Powers Tidal Disruption Events?}",
      journal = {\apj},
         year = 2015,
        month = jun,
       volume = {806},
       number = {2},
          eid = {164},
        pages = {164},
          doi = {10.1088/0004-637X/806/2/164},
archivePrefix = {arXiv},
       eprint = {1502.05792},
 primaryClass = {astro-ph.HE},
       adsurl = {https://ui.adsabs.harvard.edu/abs/2015ApJ...806..164P}
}

@ARTICLE{Shiokawa+2015,
       author = {{Shiokawa}, Hotaka and {Krolik}, Julian H. and {Cheng}, Roseanne M. and {Piran}, Tsvi and {Noble}, Scott C.},
        title = "{General Relativistic Hydrodynamic Simulation of Accretion Flow from a Stellar Tidal Disruption}",
      journal = {\apj},
         year = 2015,
        month = may,
       volume = {804},
       number = {2},
          eid = {85},
        pages = {85},
          doi = {10.1088/0004-637X/804/2/85},
archivePrefix = {arXiv},
       eprint = {1501.04365},
 primaryClass = {astro-ph.HE},
       adsurl = {https://ui.adsabs.harvard.edu/abs/2015ApJ...804...85S}
}

@software{Dastidar+2026,
  author       = {Dastidar, Ranadeep Ghosh},
  title        = {Firefly\_V2.0.0: Synchrotron Afterglow Code},
  month        = jul,
  year         = 2026,
  publisher    = {Zenodo},
  version      = {V2.0.0},
  doi          = {10.5281/zenodo.21695912},
  url          = {https://doi.org/10.5281/zenodo.21695912},
}

\end{document}